\RequirePackage{fix-cm}
\documentclass[twocolumn,epjc3]{svjour3}  
\usepackage[labelfont=bf,labelsep=space]{caption}

\smartqed  
\RequirePackage{graphicx}
\usepackage{multirow}
\usepackage[numbers,sort&compress]{natbib}
\usepackage{orcidlink}
\usepackage[fleqn]{amsmath}
\usepackage[dvips]{epsfig}
\usepackage{subcaption}
\usepackage{xcolor}
\usepackage{booktabs} 
\usepackage[T1]{fontenc}
\usepackage{microtype}
\usepackage{etoolbox}
\newcommand{\comm}[1]{}
\usepackage{hyperref}
\hypersetup{
  colorlinks=true,
  linkcolor=blue,
  citecolor=blue,
  urlcolor=blue,
  } 
\usepackage{doi}  
\journalname{Eur. Phys. J. C} 
\begin{document} 
\title{Mass-radius relationship and gravitational wave emission from  magnetized spheroidal quark stars
}
\emergencystretch=2em
\sloppy
\author{Rajasmita Sahoo\thanksref{addr1,e1}\orcidlink{0000-0002-9265-5025}
\and
Arunkarthiheyan Thiyagarajan\thanksref{addr1,e2} 
\and
Asutosh Panda\thanksref{addr1,e3}
\and
Somnath Mukhopadhyay\thanksref{addr1,e4}\orcidlink{0000-0003-2666-3200}
}
\thankstext{e1}{e-mail: \textcolor{blue}{rsphysics58@gmail.com}}
\thankstext{e2}{e-mail:  \textcolor{blue}{vel07thyag@gmail.com}}
\thankstext{e3}{e-mail: \textcolor{blue}{asutoshpanda0504@gmail.com}}
\thankstext{e4}{e-mail: \textcolor{blue}{somnath@nitt.edu} (corresponding author)}
\institute{National Institute of Technology, Tiruchirappalli, Tamil Nadu - 620015, India. 
\label{addr1}
}
\date{Received: date / Accepted: date}
\maketitle
\begin{abstract} 
In this work, we investigate the structure and gravitational wave (GW) signatures of strongly magnetized, oblate spheroidal quark stars by employing an anisotropic equation of state (EoS) derived from the MIT Bag model, extended to include the effects of density-dependent strong magnetic fields and the resulting pressure anisotropy arising from the breaking of spatial symmetry. Both magnetized strange quark matter (MSQM) and magnetized color-flavor locked (MCFL) phases are examined within the framework of the $\gamma$-metric formalism, which captures the deviation from spherical symmetry. We compute the mass-radius relation, ellipticity, gravitational redshift, mass quadrupole moment and tidal deformability for representative bag constants of $\rm{65\,MeV/fm^3}$ and $\rm{75\,MeV/fm^3}$. Using the obtained quadrupole moments, we further estimate the continuous gravitational wave strain amplitude ($h_{0}$) for isolated deformed rotating quark stars. Our results indicate that density-dependent strong magnetic fields and color superconductivity can significantly alter stellar compactness and yield gravitational wave signals, potentially detectable by next generation observatories like the Einstein Telescope and Cosmic Explorer.
\end{abstract}
\section{Introduction}
\label{intro}
Compact stars serve as natural laboratories for probing the behavior of matter under the most extreme conditions of density, pressure, and magnetic field strength. Among the various classes of such compact remnants, strange quark stars (SQSs)--hypothetical stars composed entirely of deconfined up (${u}$), down (${d}$) and strange (${s}$) quarks--offer a unique opportunity to study matter governed by strong interactions. The concept was first introduced by Bodmer and later developed by Witten~\cite{Bo71, Wi84}, who proposed that strange quark matter (SQM) may represent the absolute ground state of hadronic matter at zero pressure, provided that the baryon density exceeds a critical threshold. This transition from hadronic to deconfined quark matter constitutes a first order phase transition predicted by quantum chromodynamics (QCD) under extreme conditions~\cite{Iw04}. Two distinct regimes are associated with this transition: one at high temperature and low baryon chemical potential, relevant to the early universe and another at low temperature and high baryon density, typical of compact stellar interiors~\cite{Bo11}. Under such circumstances, quarks become deconfined and weak interactions generate strangeness, producing a chemically equilibrated mixture of $u$, ${d}$, and ${s}$ quarks. If the conversion extends throughout the star, the result is a self bound strange quark star~\cite{It70,Al86}. These stars are predicted to be extremely compact with radii of $\rm{\sim 8\, km}$, and masses around $\rm{1.5\,M_{\odot}}$ and can potentially explain several compact objects whose properties challenge standard neutron star models~\cite{Pr02, Abu08, Gango13, Ra11}. However, the softness of many quark matter equations of state (EoSs) makes it difficult to reproduce recently measured massive pulsars, such as $\rm{PSR\ J0348+0432}$ $\rm{(2.01 \pm  0.04)\,M_{\odot}}$, $\rm{PSR\ J0740+6620}$ $\rm{(2.08 \pm 0.07)\, M_{\odot}}$, and $\rm{PSR\ J0952-0607}$ $\rm{(2.35\pm 0.17)\, M_{\odot}}$~\cite{Antoni13, Mil21, Li18,Ro22}. This discrepancy has motivated the development of stiffer EoSs, incorporating mechanisms such as quark interactions or superconducting pairing to sustain higher-mass configurations~\cite{Al03, Bal03,Lai11,Bona12, Chu17}.
A further crucial factor influencing compact star structure is the presence of ultra strong magnetic fields. Magnetars, for instance, exhibit surface magnetic fields of the order of $\rm{10^{15}\,G}$, while the core fields may exceed $\rm{5\times 10^{18}\,G}$~\cite{Mi90,Tho93}. Such intense fields alter the EoS through Landau quantization and generate pressure anisotropy, since the stress-energy tensor becomes direction dependent. Consequently, the perpendicular and parallel components of the pressure (${P_{\perp}}$ and ${P_{\parallel}}$) differ, breaking spherical symmetry and inducing axisymmetric deformation~\cite{Fer10, Bali14,Kon99, Deba15, Ter19, Sahoo24}. Earlier works have examined magnetized strange quark stars under the assumption of a uniform magnetic field throughout the stellar interior~\cite{Terr21}. While this approximation simplifies the analysis, it neglects the natural spatial variation of magnetic fields expected in realistic compact stars. Moreover, models based solely on magnetized strange quark matter (MSQM) may underestimate the stiffness of the EoS at high densities. At such densities, quark pairing becomes energetically favorable, leading to the color-flavor locked (CFL) phase---a color superconducting state in which all quarks form Cooper pairs~\cite{Alf99,alf98,alf08}. This phase allows the existence of a rotated photon field ${\tilde{A}_{\mu}}$, resulting in a modified magnetic response distinct from that of ordinary superconductors~\cite{Alf00,Gor00, Fuk08}. A physically consistent treatment of the CFL phase also requires electromagnetic and color charge neutrality to ensure thermodynamic stability~\cite{Alfo02,Raj01,Buba05,go11}. Recent studies incorporating magnetic fields into the CFL framework~\cite{go11, Man15, Goswa23} have shown that the magnetized color-flavor locked (MCFL) phase can produce more massive configurations than the unpaired MSQM phase, highlighting the crucial role of pairing and anisotropy. However, these models typically assume constant magnetic fields and spherical symmetry, limiting their applicability to strongly deformed and realistic stars. In this work, we address these limitations by modeling magnetized spheroidal quark stars using a density-dependent magnetic field profile and an anisotropic EoS that captures the effects of pressure anisotropy. Both MSQM and MCFL phases are considered within the ${\gamma}$-metric formalism, which allows for departures from spherical symmetry. We compute key stellar properties, including the mass-radius relation, ellipticity, gravitational redshift, mass quadrupole moment, tidal deformability, and gravitational wave strain amplitude for representative bag constants of ${B_{bag}}=\rm{65\,MeV/fm^3}$ and $\rm{75\,MeV/fm^3}$, and for pairing gaps of ${\Delta=\rm{30\,MeV}}$ and $\rm{50\,MeV}$ in the MCFL phase. By extending the traditional isotropic TOV approach to include anisotropic pressure and deformation effects, this study provides a more realistic framework for exploring the structure and potential gravitational wave emission of magnetized quark stars. The paper is organized as follows. In Section~\ref{sec:1} we present the anisotropic equation of state (EoS) for magnetized strange quark matter, emphasizing Landau quantization and magnetic pressure anisotropy. Section~\ref{sec:2} extends the formulation to the magnetized color-flavor locked (MCFL) phase, incorporating superconducting pairing and neutrality constraints. In Section~\ref{sec:3}, we discuss stability criteria for anisotropic quark matter under strong magnetic fields. Section~\ref{sec:4} describes the stellar structure framework based on the ${\gamma}$-metric and the modified equilibrium equations for deformed, axisymmetric configurations. In Section~\ref{sec:5}, we introduce density-dependent magnetic field profile and its implementation. Section~\ref{sec:6} presents the results---mass-radius relations, deformation measures (ellipticity and quadrupole moment), gravitational redshift, tidal deformability, and estimates of the continuous gravitational wave strain amplitude---with emphasis on the roles of magnetic-field anisotropy and pairing. Finally, Section~\ref{sec:7} summarizes our conclusions and astrophysical implications.
\section{Anisotropic Equation of State of Magnetized Strange Quark Matter (MSQM)}
\label{sec:1}
In this section, we formulate the anisotropic equation of state (EoS) for magnetized strange quark matter (MSQM), composed of up (${u}$), down (${d}$) and strange (${s}$) quarks along with electrons, in the presence of strong magnetic field. The system is assumed to be in ${\beta}$-equilibrium and electrically charge neutral. The magnetic field ${B}$ is taken to be aligned along the z direction i.e., ${(B = (0,0,B))}$. Under such conditions, the motion of charged particles becomes quantized in the plane perpendicular to the magnetic field, forming discrete Landau levels. We employ the phenomenological MIT Bag model, which represents quarks as quasi-free particles confined within a finite region (the "bag") by a vacuum pressure termed as the bag constant ${B_{bag}}$. This constant effectively accounts for color confinement and modifies both the energy density and pressure of the system~\cite{Chodos74}. The quark masses and electric charges are taken as ${m_{u}= \rm{2.2\, MeV}}$, ${m_{d}= \rm{4.8\, MeV}}$, ${m_{s}=\rm{95\, MeV}}$ with corresponding charges ${e_{u} = + \rm{\frac{2}{3} e}}$ and ${e_{d} = e_{s}= -\rm{\frac{1}{3} e}}$. The electron mass is ${m_{e} = \rm{0.511\; MeV}}$. The single particle energy of a charged fermion occupying the $\nu^{th}$ Landau level is given by~\cite{Strick12,Mukho15,Sahoo24}:
\begin{equation}\label{eq:Energy}
E_{\nu,p_{z}}^{(i)}=[{p_{z}^2 c^2}+m_{i}^2 c^4(1+2\nu B_{D}^{(i)})]^{1/2}.
\end{equation}
where ${i \in [u,d,s,e]}$ denotes the particle species, ${p_{z}}$ is the component of the momentum along the magnetic field, and ${m_{i}}$ is the rest mass of the $i^{th}$ particle species. The Landau level index is defined as ${\nu = n +\frac{1}{2}+s_{z}}$, where ${n = 0,1,2,...}$ is the Landau levels and ${s_{z} = \pm \frac{1}{2}}$ is the spin projection along the field direction. The dimensionless magnetic field ${B_{D}^{(i)} = \frac{B}{B_{c}^{(i)}}}$, where ${B}$ is the magnetic field strength and ${B_{c}^{(i)} = \frac{m_{i}^2 c^3}{e_{i}\hbar}}$ is the critical magnetic field for each species.
The density of states (including spin degeneracy) for each particle species ${i \in [u,d,s,e]}$ is given by,
\begin{equation}\label{eq:dos}
{\sum_{\nu} \frac{2\pi}{h^2} m_{i}^2c^2 B_{D}^{(i)} g_{\nu} \int\limits \frac{dp_{z}}{h}}.
\end{equation}
where ${g_{\nu}}$ is the spin degeneracy factor, with ${g_{\nu} = 1}$ for ${\nu =0}$ and $\rm{g_\nu = 2}$ for ${\nu \ge 1}$. At absolute zero temperature, the Fermi-Dirac distribution function simplifies to,
\begin{equation}
{f(E)=}
\begin{cases}
    1, \; \mbox{for } {E\leq E_{F}^{(i)}} \\
    0, \; \mbox{for } {E> E_{F}^{(i)}}
\end{cases}
\end{equation}
where ${E_{F}^{(i)}}$ is the Fermi energy of the species ${i}$, defining the maximum occupied energy level. This condition imposes a constraint on ${p_{z}}$ such that:
\begin{equation}\label{eq:pzf}
{p_{z,F}^{(i)}(\nu)= \frac{1}{c}\sqrt{{\left(E_{F}^{(i)}\right)}^2-m_{i}^2 c^4\left(1+2\nu B_{D}^{(i)}\right)}}.
\end{equation}
For each species, the Landau level index is constrained by the condition,
${\nu\leq\nu_{max}^{(i)}=\dfrac{1}{2 B_D^{(i)}}\left(\dfrac{{\left(E_F^{(i)}\right)}^2}{m_{i}^2 c^4}-1\right)}$
ensuring that only a finite number of Landau levels are populated at a given magnetic field strength. 
Since magnetized strange quark matter (MSQM) consists of fluid of quarks and electrons in the presence of strong magnetic field, the equation of state (EoS) for MSQM can be written as derived in~\cite{Strick12,Mukho15,Sahoo24}. The number density ${n_{i}}$ of any charged fermion species ${i}$ (quark or electron) in MSQM phase as,
\begin{eqnarray}\label{eq:nden}
{n_i} &=& {\dfrac{2\pi}{h^2}}\, {f_{i}}\, {m_i^2 c^2 B_D^{(i)}} {\sum_{\nu=0}^{\nu_{\max}^{(i)}} }{g_{\nu}}{\int_{-\infty}^{+\infty} f(E)}\, {\frac{dp_{z}}{h}}, \nonumber \\
&=& {\dfrac{4\pi}{h^2}}\, {f_{i}}\, {m_i^2 c^2 B_D^{(i)}}
{\sum_{\nu=0}^{\nu_{\max}^{(i)}}} {g_{\nu}}
{\int_{0}^{p_{z,F}^{(i)}(\nu)} \frac{dp_{z}}{h}}, \nonumber \\
&=& {\dfrac{4\pi}{h^3}}\, {f_{i}}\, {m_i^2 c^2 B_D^{(i)}}
{\sum_{\nu=0}^{\nu_{\max}^{(i)}} g_{\nu}}\,
{p_{z,F}^{(i)}(\nu)}.
\end{eqnarray}
where, ${f_{i}}$ is denoted as the flavor degeneracy factor. For each quark, ${f_{u,d,s}=3}$ and for electron, ${f_{e}=1}$.
The energy density ${\epsilon_{i}}$ of each particle species (quark or electron) in the presence of magnetic field in MSQM phase is given as, 
\begin{eqnarray}\label{eq:enden}
{\epsilon_i} &=& {\dfrac{2\pi}{h^2}}\, {f_{i}}\, {m_{i}^2 c^2 B_D^{(i)} }
{\sum_{\nu=0}^{\nu_{\max}^{(i)}}} {g_{\nu}} 
{\int\limits_{-\infty}^{+\infty} f(E)}\, {E_{\nu,p_z}^{(i)}} {\frac{dp_{z}}{h}}, \nonumber\\
&=& {\dfrac{4\pi}{h^2}}\, {f_{i}}\, {m_{i}^2 c^2 B_D^{(i)}} 
{\sum_{\nu=0}^{\nu_{\max}^{(i)}}} {g_{\nu} }
{\int\limits_{0}^{p_{z,F}^{(i)}(\nu)} 
\sqrt{p_{z}^2 c^2 + m_{i}^2 c^4(1+2\nu B_{D}^{(i)})}}\,
{\frac{dp_{z}}{h}}, \nonumber\\
&=& {\dfrac{2\pi}{h^3}}\, {f_{i}}\, {m_{i}^2 c^2 B_D^{(i)}} 
{\sum_{\nu=0}^{\nu_{\max}^{(i)}}} {g_{\nu}} 
{\Bigg[ E_F^{(i)}\, p_{z,F}^{(i)}(\nu) 
+ m_{i}^2 c^3 (1+2\nu B_{D}^{(i)})} \nonumber\\
&& {\times \log_e \!\left( 
\frac{E_F^{(i)} + p_{z,F}^{(i)}(\nu)c}
{\sqrt{m_{i}^2 c^4(1+2\nu B_{D}^{(i)})}}
\right) \Bigg]}.
\label{eq:enden}
\end{eqnarray}
The parallel pressure (z-direction) of each particle species (quark or electron) in the direction of the magnetic field in MSQM phase is given by,
\begin{eqnarray}\label{eq:ppar}
{P_{\parallel,i}} &=& {f_{i}}\, {B_D^{(i)}}\,{\frac{2\pi m_i^2 c^2}{h^3}}\,
{\sum_{\nu=0}^{\nu_{\max}^{(i)}}} {g_{\nu}}
{\int_{-\infty}^{+\infty} 
\frac{c^2 p_{z}^2}{E_{\nu,p_z}^{(i)}}}\, {f(E)}\, {dp_{z}}, \nonumber \\
&=& {f_{i}}\, {B_D^{(i)}}\, {\frac{4\pi m_i^2 c^2}{h^3}}\,
{\sum_{\nu=0}^{\nu_{\max}^{(i)}}} {g_{\nu}}
{\int\limits_{0}^{p_{z,F}^{(i)}(\nu)}
\frac{c^2 p_{z}^2}
{\sqrt{p_{z}^2 c^2 + m_i^2 c^4 (1 + 2\nu B_D^{(i)})}}}\, {dp_{z}}, \nonumber \\
&=& {f_{i}}\, {\frac{2\pi m_i^2 c^2 B_D^{(i)}}{h^3}}\,
{\sum_{\nu=0}^{\nu_{\max}^{(i)}}} {g_{\nu}}
{\Bigg[
E_F^{(i)}\, p_{z,F}^{(i)}(\nu)
- m_i^2 c^3 (1 + 2\nu B_D^{(i)})} \nonumber \\
&& {\times \log_{e}\!\left(
\frac{E_F^{(i)} + p_{z,F}^{(i)}(\nu)c}{
\sqrt{m_i^2 c^4 (1 + 2\nu B_D^{(i)})}}
\right)
\Bigg]}.
\end{eqnarray}
The perpendicular pressure (x- and y- directions) of each particle species (quark or electron) in the direction perpendicular to the magnetic field in MSQM phase is given by,
\begin{eqnarray}
\label{eq:pperp}{P_{\perp,i}} &=& {f_{i}}\,\rm{\frac{m_i^2 c^4}{2}}\,{\left(B_D^{(i)}\right)^2
\left(\frac{2\pi m_i^2 c^2}{h^3}\right)}
{\sum_{\nu=0}^{\nu_{\max}^{(i)}} g_{\nu} \times} \nonumber \\  
&& {\int_{-\infty}^{+\infty} 
\frac{2\nu}{E_{\nu,p_z}^{(i)}}}\, {f(E)}\, {dp_{z}}, \nonumber \\
&=& {f_{i}}\, {m_i^2 c^4 \left(B_D^{(i)}\right)^2
\left(\frac{2\pi m_i^2 c^2}{h^3}\right)}
{\sum_{\nu=0}^{\nu_{\max}^{(i)}}} {g_{\nu} \times} \nonumber \\
&& {\int\limits_{0}^{p_{z,F}^{(i)}(\nu)} 
\frac{2\nu}{\sqrt{p_{z}^2 c^2 + m_i^2 c^4(1 + 2\nu B_D^{(i)})}}}\, {dp_{z}}, \nonumber \\
&=& {f_{i}}\, {m_i^2 c^3 \left(B_D^{(i)}\right)^2
\left(\frac{2\pi m_i^2 c^2}{h^3}\right)
\sum_{\nu=0}^{\nu_{\max}^{(i)}} g_{\nu} \times}  \nonumber \\
&& {2\nu}\, {\log_{e}\!\left(
\frac{E_F^{(i)} + p_{z,F}^{(i)}(\nu)c}{
\sqrt{m_i^2 c^4 (1 + 2\nu B_D^{(i)})}}
\right)}.
\end{eqnarray}
These expressions explicitly demonstrate the anisotropy induced by the magnetic field, with ${P_{\perp,i}<P_{\parallel,i}}$ at high ${B}$. Physically, this arises because motion perpendicular to the field is quantized into Landau levels, restricting transverse momentum and reducing the corresponding pressure component.
The system satisfies the conditions of ${\beta}$-equilibrium, charge neutrality, and total baryon number density conservation~\cite{Feli08};\\
\begin{eqnarray}
\label{eq:nb}
{\mu_{u}+ \mu_{e} = \mu_{d}= \mu_{s}}, \nonumber\\
{\frac{2}{3} n_{u}-\frac{1}{3}(n_{d}+n_{s})-n_{e} = 0}, \nonumber\\
{n_{B}=\frac{1}{3}(n_{u}+n_{d}+n_{s})}.
\end{eqnarray}
where ${\mu_{i}}$ is the chemical potential of species ${i \in [u,d,s,e]}$. We neglect neutrino trapping, assuming that they escape freely implying they have no influence on the ${\beta}$-equilibrium condition. For a given baryon number density ${n_{B}}$ and a specified central magnetic field strength $B_{cen}$, these equations along with Eqs.~\eqref{eq:nden}  can be solved to obtain the chemical potentials and compute all the thermodynamic quantities of the system. To construct the EoS for MSQM phase, we adopt the MIT Bag Model. In this model, the equation of state includes the bag constant (representing the vacuum energy) and also incorporates the magnetic field contributions. Specifically, the magnetic field adds an energy density term ${\epsilon_{B}=\frac{B^2}{8\pi}}$, along with anisotropic pressure components: ${P_{\perp, B}=\frac{B^2}{8\pi}}$ and ${P_{\parallel, B}= -\frac{B^2}{8\pi}}$, which arise due to the presence of the magnetic field anisotropy itself~\cite{Das12,Sahoo24}.
\begin{equation}
{\epsilon_{T}^{MSQM}=\sum_{i=u,d,s,e}\epsilon_{i}+B_{bag}+\frac{B^2}{8 \pi}},\label{eq:et1}
\end{equation}
\begin{equation} 
\label{eq:pr1}
{P_{\parallel T}^{MSQM}=\sum_{i=u,d,s,e}P_{\parallel, i}-B_{bag}-\frac{B^2}{8 \pi}},
\end{equation}
\begin{equation} 
\label{eq:pn1}
{P_{\perp T}^{MSQM}=\sum_{i=u,d,s,e}P_{\perp, i}-B_{bag}+\frac{B^2}{8 \pi}}.
\end{equation}
The contrasting signs of the magnetic terms in ${P_{\parallel T}^{MSQM}}$ and ${P_{\perp T}^{MSQM}}$ reflect the intrinsic magnetic pressure anisotropy, which causes the stellar configuration to become oblate under strong magnetic fields.
\section{Anisotropic Equation of State of Magnetized Color-Flavor Locked (MCFL) Quark Matter}
\label{sec:2}
In the magnetized color-flavor locked (MCFL) phase, quarks form Cooper pairs across different flavors (${u,d,s}$) due to the phenomenon of color superconductivity. The baryon number density ${n_{B}}$ is assumed to be equal for all quark flavors, modified by the pairing energy contribution that stabilizes the superconducting state. This condition ensures both electric and color neutrality, eliminating the need for electrons in this phase. The charge neutrality condition for MCFL matter can be expressed as~\cite{go11,Man15}:
\begin{equation}\label{eq:nbcfl}
{n_{u}+\frac{2\Delta^2 \mu_B}{\pi^2}=n_{d}+\frac{2\Delta^2 \mu_B}{\pi^2}=n_{s}+\frac{2\Delta^2 \mu_B}{\pi^2}=n_{B}}.
\end{equation}
where ${n_{i}=n_{u,d,s}}$ are the number densities obtained from Eqs.~\eqref{eq:nden}, ${\mu_{B}}$ is the baryon chemical potential, and ${\Delta}$ is the pairing gap parameter. The additional term ${\left(\frac{2\Delta^2 \mu_B}{\pi^2}\right)}$ arises due to quark pairing energy in the MCFL phase, effectively reducing the total energy of the system and favoring the superconducting state. \\
The total energy density and anisotropic pressure components (parallel and perpendicular to the magnetic field) for MCFL quark matter are written as~\cite{go11,Man15,Raja01}:
\begin{equation}
\label{eq:et}
{\epsilon_{T}^{MCFL}=\sum_{i=u,d,s}\epsilon_{i}+B_{bag}+\frac{B^2}{8 \pi}-\frac{3\Delta^2\mu_{B}^2}{\pi^2}},
\end{equation}
\begin{equation} 
\label{eq:pr}
{P_{\parallel T}^{MCFL}=\sum_{i=u,d,s}P_{\parallel, i}-B_{bag}-\frac{B^2}{8 \pi}+\frac{3\Delta^2\mu_{B}^2}{\pi^2}},
\end{equation}
\begin{equation} 
\label{eq:pn}
{P_{\perp T}^{MCFL}=\sum_{i=u,d,s}P_{\perp, i}-B_{bag}+\frac{B^2}{8 \pi}+\frac{3\Delta^2\mu_{B}^2}{\pi^2}}.
\end{equation}
for ${i \in [u,d,s]}$, the terms ${\epsilon_{i}}$, ${P_{\parallel, i}}$ and ${P_{\perp,i}}$ represent the contributions from individual unpaired quarks (same as in the MSQM phase). The bag constant ${B_{bag}}$ accounts for the confinement energy, while the magnetic field introduces an anisotropic contribution to the total pressure, increasing the perpendicular pressure (${P_{\perp}}$) and reducing the parallel pressure (${P_{\parallel}}$). This anisotropy arises from the magnetic stress tensor, which breaks the spatial isotropy and deforms the star into an oblate configuration. The final terms proportional to ${\left({\Delta^2\mu_{B}^2}\right)}$, correspond to the pairing energy that lowers the free energy of the system. Physically, this term enhances the stability of the quark matter against gravitational collapse leading to denser and more compact configurations compared to the unpaired MSQM phase. In this work, we adopt two representative gap parameters, ${\Delta=\rm{30\,MeV}}$ and $\rm{{50\,MeV}}$, assuming that the pairing gap is independent of magnetic field strength. These values capture the possible range of superconducting effects in magnetized quark matter and their influence on the star's structure and gravitational wave emission.
\begin{figure*}[t]
\centering
\begin{subfigure}[b]{0.45\textwidth}
\includegraphics[width=\linewidth]{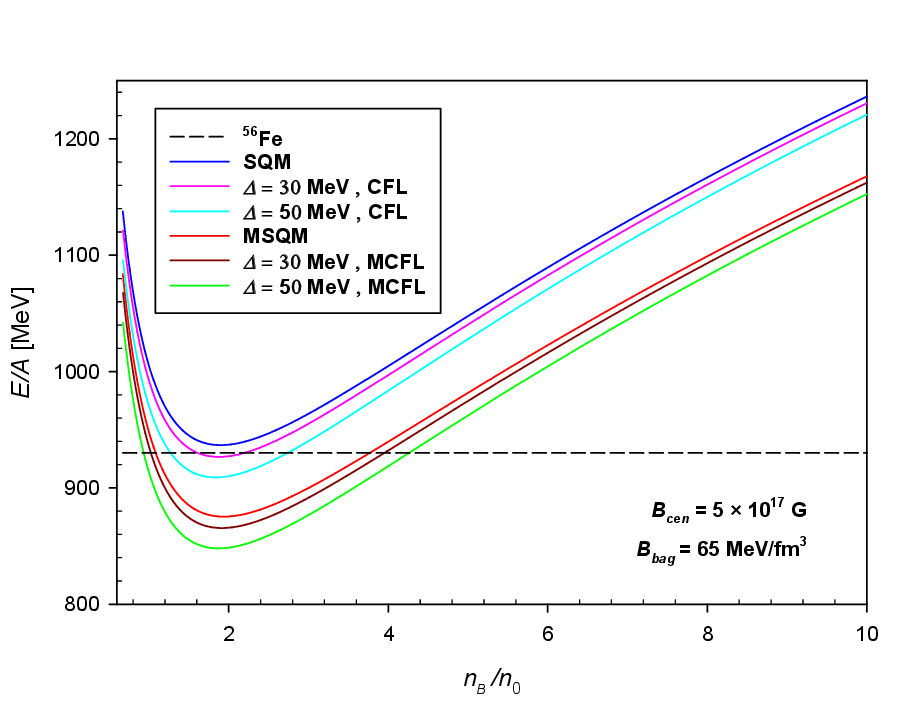}
\end{subfigure}
\hfill
\begin{subfigure}[b]{0.45\textwidth}
\includegraphics[width=\linewidth]{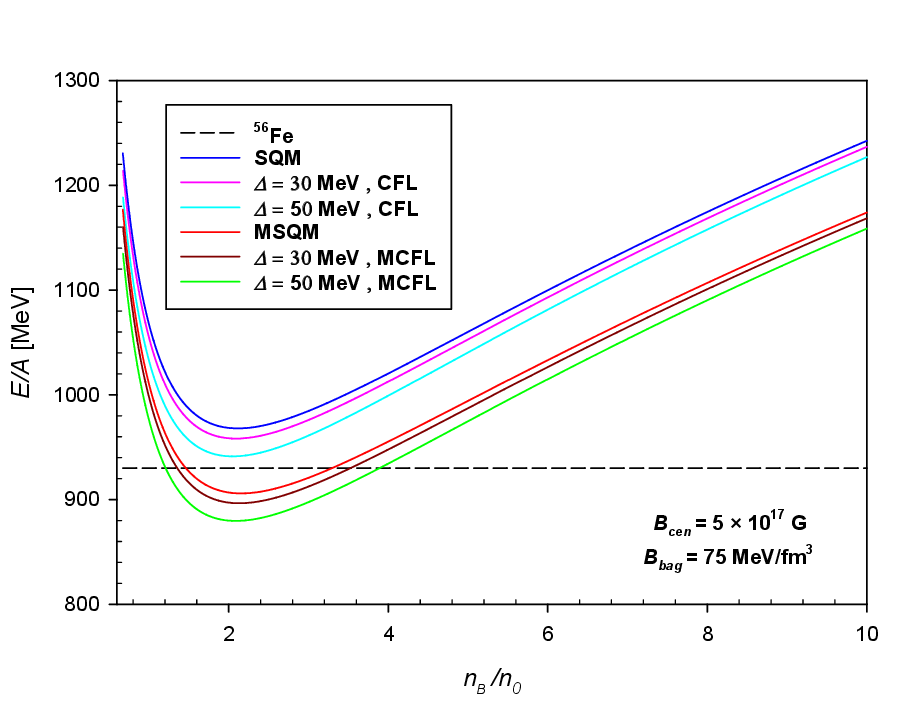}
\end{subfigure}
\vspace{0.1cm}  
\begin{subfigure}[b]{0.45\textwidth}
\includegraphics[width=\linewidth]{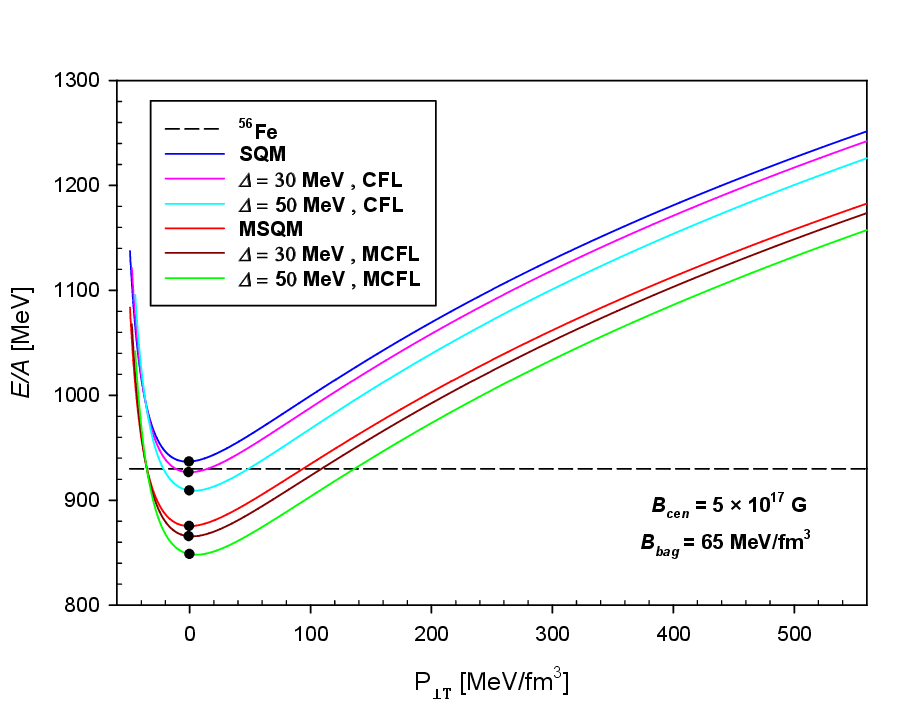}
\end{subfigure}
\hfill
\begin{subfigure}[b]{0.45\textwidth}
\includegraphics[width=\linewidth]{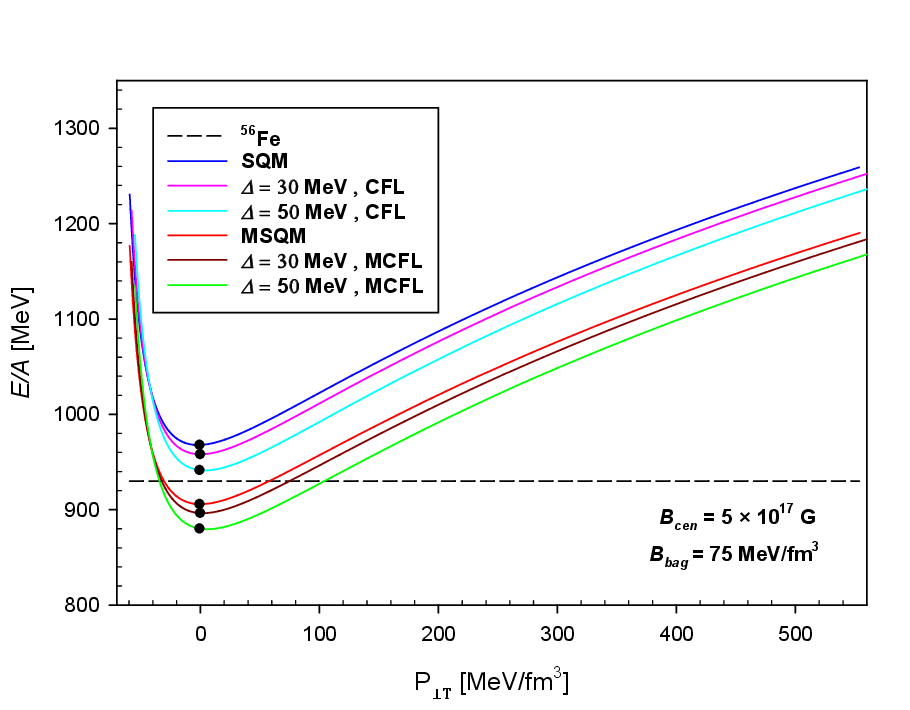}
\end{subfigure}
\caption{Energy per baryon $E/A$ as a function of the baryon density ratio $n_{B}/n_{0}$ and total perpendicular pressure $P_{\perp T}$ for a central magnetic field of $\rm{5 \times 10^{17}}\,G$. The upper row panels correspond to $B_{bag}=\rm{65\,MeV/fm^{3}}$ and $B_{bag}=\rm{75\,MeV/fm^{3}}$, respectively. The upper row panels show the variation of $E/A$ with the baryon density ratio $n_{B}/n_{0}$, while the lower row panels depict the corresponding variation with the total perpendicular pressure $P_{\perp T}$ for the same bag constants. Results are shown for different phases of quark matter, namely strange quark matter (SQM), color-flavor locked (CFL) matter with pairing gaps $\Delta=\rm{30\,MeV}$ and $\Delta=\rm{50\,MeV}$, magnetized strange quark matter (MSQM), and magnetized color-flavor locked (MCFL) matter with the same pairing gaps. The horizontal dashed line denotes the energy per baryon of the ${^{56}{Fe}}$ nucleus, and the black dots mark the points where the total perpendicular pressure vanishes. }
\label{fig:stability_plot}
\end{figure*}
\begin{figure*}[t]
\centering
\begin{subfigure}[b]{0.45\textwidth}
\includegraphics[width=\linewidth]{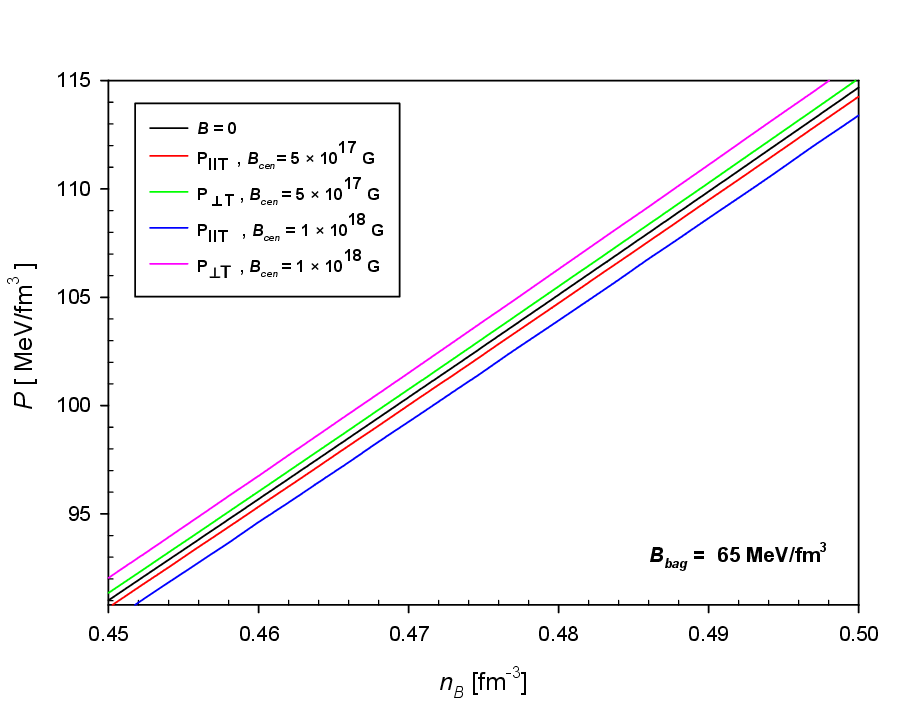}
\end{subfigure}
\hfill
\begin{subfigure}[b]{0.45\textwidth}
\includegraphics[width=\linewidth]{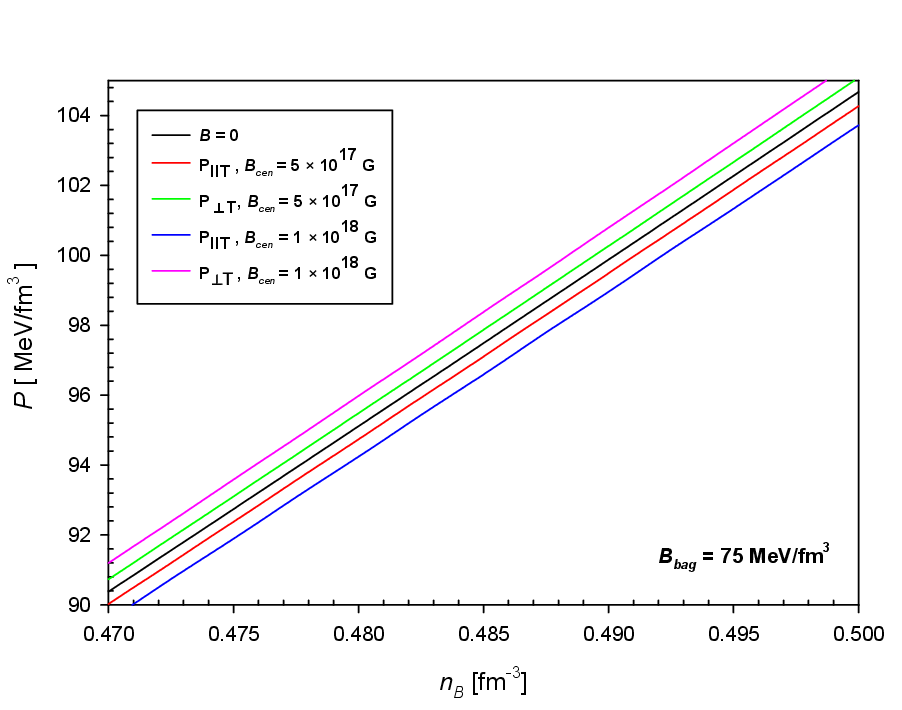}
\end{subfigure}
\vspace{0.1cm}  
\begin{subfigure}[b]{0.45\textwidth}
\includegraphics[width=\linewidth]{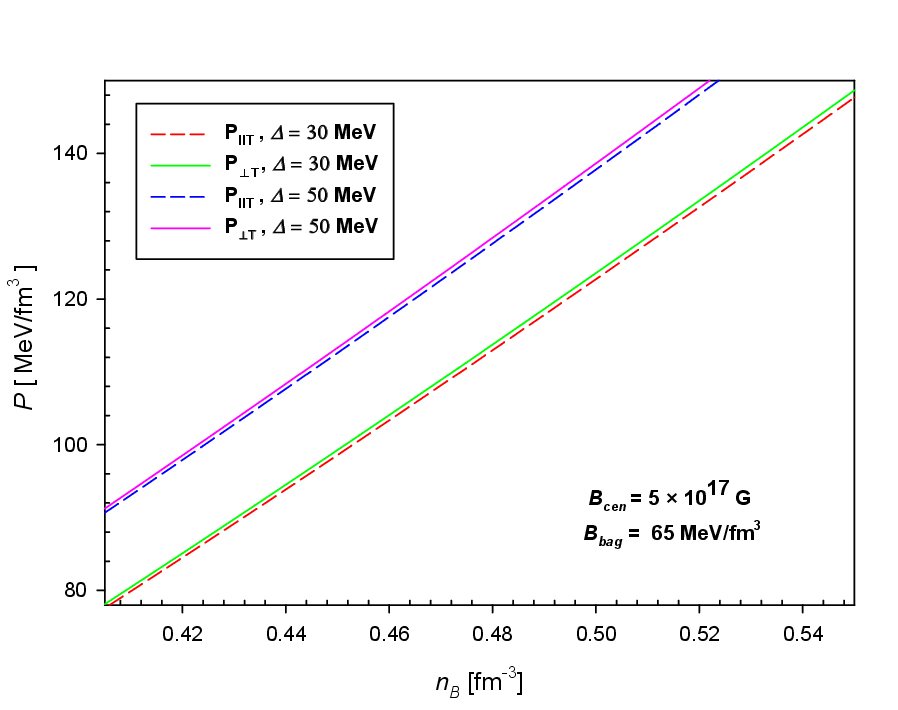}
\end{subfigure}
\hfill
\begin{subfigure}[b]{0.45\textwidth}
\includegraphics[width=\linewidth]{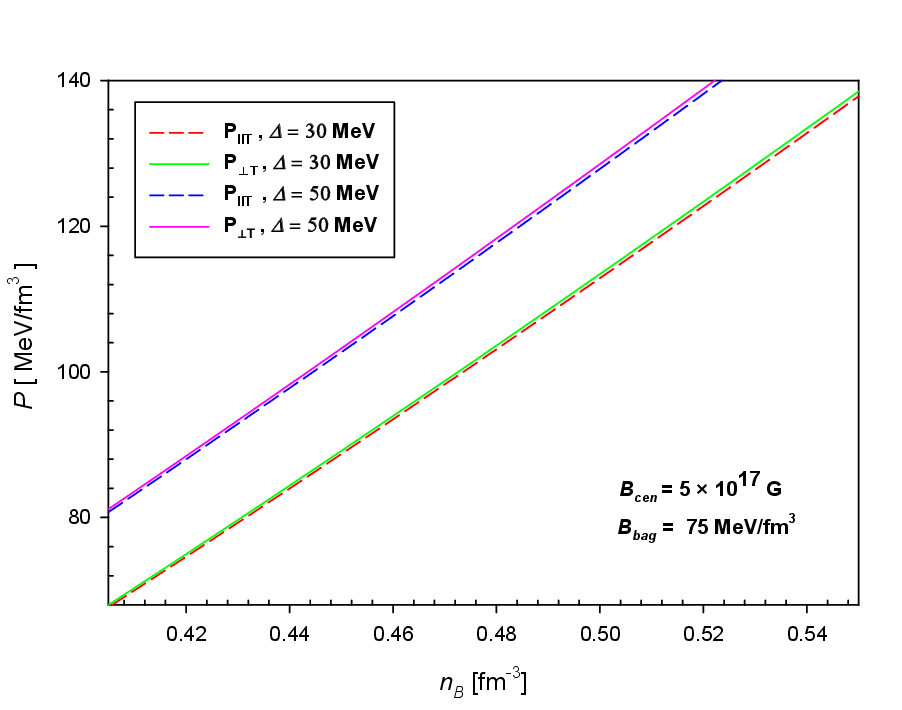}
\end{subfigure}
\caption{Variation of pressure $P$ with baryon number density $n_{B}$ under different physical conditions. The upper row panels correspond to the magnetized strange quark matter (MSQM) phase and show the behavior of the total parallel pressure $P_{\parallel T}$ and total perpendicular pressure $P_{\perp T}$ for central magnetic field strengths $B_{cen}=\rm{5 \times 10^{17}\,G}$ and $\rm{1\times 10^{18}\,G}$ at fixed bag constants $B_{bag}=\rm{65\,MeV/fm^3}$ and $\rm{75\,MeV/fm^3}$, respectively. The black solid line represents the non-magnetized case ($B=0$). The lower row panels correspond to the magnetized color-flavor locked (MCFL) phase and illustrate the effect of different pairing gaps, $\Delta=\rm{30\,MeV}$ and $\Delta=\rm{50\,MeV}$, on both $P_{\parallel T}$ and $P_{\perp T}$ for the same bag constants at a central magnetic field strength of $B_{cen}=\rm{5\times 10^{17}}\,G$.}
\label{fig:press_nb_plot}
\end{figure*}
\section{Stability Condition for Anisotropic Magnetized Strange Quark Matter}
\label{sec:3}
\begin{figure}[t]
  \vspace{0.0cm}
\eject\centerline{\epsfig{file= 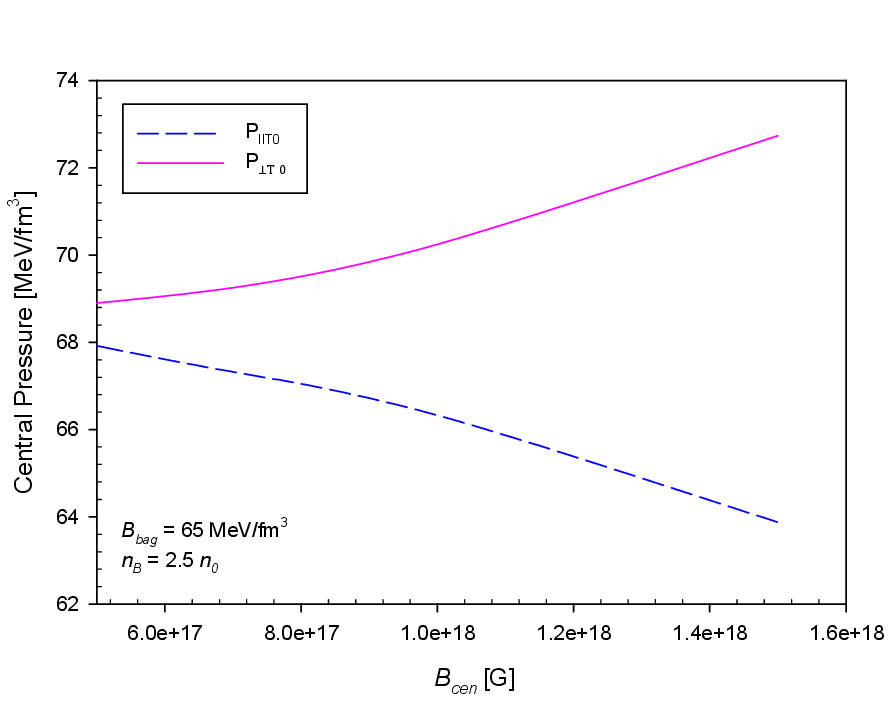,height=6.5cm,width=8.5cm}}
 \caption{Variation of central pressure with central magnetic field strength for a fixed central baryon number density $n_{B}=2.5\, n_{0}$ and bag constant $B_{bag}=\rm{65 \; MeV/fm^{3}}$.}
 \label{fig:PBdc}
\end{figure}

The stability of magnetized strange quark matter and its color-superconducting phases can be assessed by comparing the energy per baryon (${E/A}$) at zero pressure with that of the most stable atomic nucleus ${^{56}{Fe}}$, whose energy per baryon is about $\rm{930\,MeV}$~\cite{pau08,wen13,go11}. The stability criterion is thus expressed as,
\begin{equation}
{\left(\frac{E}{A}\right)_{P=0}<\left(\frac{E}{A}\right)_{^{56}\mathrm{Fe}} \simeq\, 930 \,MeV}. 
\end{equation}
If this inequality is satisfied, the quark matter phase is absolutely stable otherwise it remains metastable. The upper row panels of Fig.~\hyperref[fig:stability_plot]{\ref*{fig:stability_plot}} show the variation of ${E/A}$ with the normalized  baryon density (${n_{B}/n_{0}}$), where $n_{0}=\rm{0.16\,fm^{-3}}$ denotes the nuclear saturation density. For $B_{bag}=\rm{65\,MeV/fm^3}$, the MSQM, CFL, and MCFL phases with pairing gaps $\Delta=\rm{30\,MeV}$ and $\rm{50\,MeV}$ lie close to or even below the ${^{56}{Fe}}$ line, indicating potential absolute stability. When the bag constant is increased to $B_{bag}=\rm{75\,MeV/fm^3}$, the curves shift upward, suggesting a reduction in stability; however, the MSQM and MCFL phases with higher pairing gaps remain comparatively stable. The non-magnetized SQM and CFL phases by contrast lie well above the stability limit, signifying metastability. A similar trend is evident in the lower row panels of Fig.~\hyperref[fig:stability_plot]{\ref*{fig:stability_plot}}, where ${E/A}$ is plotted as a function of the total perpendicular pressure ${P_{\perp T}}$. The zero pressure points mark the equilibrium configurations that determine absolute stability. For $B_{bag}=\rm{65\,MeV/fm^3}$, the MCFL phase with $\Delta=\rm{50\,MeV}$ lies notably below the ${^{56}{Fe}}$ threshold, confirming its absolute stability while higher bag constant weaken this effect.
The overall stability of magnetized quark matter depends sensitively on the bag constant, quark masses, pairing gap, and magnetic field strength. In the present analysis, the quark masses and pairing gaps are taken to be relatively small, so that the SQM and CFL phases do not fully satisfy the stability condition. Increasing these parameters reduces the total free energy per baryon, allowing even these phases to achieve absolute stability.
The presence of strong magnetic field further enhances stability by Landau quantization which lowers the transverse kinetic energy and thus the total energy per baryon. This effect explains why the MSQM and MCFL phases emerge as the most stable configurations under intense fields. In Fig.~\hyperref[fig:press_nb_plot]{\ref*{fig:press_nb_plot}}, we show the relation between pressure and baryon number density for the MSQM and MCFL phases at different central magnetic fields. The upper row panels of Fig.~\hyperref[fig:press_nb_plot]{\ref*{fig:press_nb_plot}} (for $B_{bag}=\rm{65\,MeV/fm^3}$ and $\rm{75\,MeV/fm^3}$) reveal a clear pressure anisotropy induced by the magnetic field, where ${P_{\perp T}>P_{\parallel T}}$. The anisotropy diminishes at higher bag constants as the confining vacuum pressure dominates over the magnetic contribution. Fig.~\hyperref[fig:PBdc]{\ref*{fig:PBdc}} quantifies this anisotropy showing that the difference between ${P_{\perp T}}$ and ${P_{\parallel T}}$ increases with field strength, a hallmark of deformation in magnetized quark stars. The lower row panels of Fig.~\hyperref[fig:press_nb_plot]{\ref*{fig:press_nb_plot}} illustrate the effect of pairing on the MCFL phase. Increasing the gap from $\Delta=\rm{30\,MeV}$ to $\rm{50\,MeV}$ stiffens the equation of state leading to higher pressures at a given baryon number density. Both strong magnetic fields and larger pairing gaps thus act to stiffen the EoS, enhance mechanical stability and favor the formation of massive, oblate stellar configurations in strong magnetic environments.
\section{Stellar Structure Equations for Anisotropic Magnetized Quark Stars}
\label{sec:4}
In the presence of strong magnetic field, the pressure inside a compact star becomes anisotropic i.e., the pressure along the magnetic field direction (${\parallel}$) differs from that perpendicular to it (${\perp}$). This anisotropy breaks spherical symmetry and causes the star to assume an oblate axisymmetric configuration. To describe the structure of such magnetically deformed quark stars both for magnetized strange quark matter (MSQM) and magnetized color-flavor locked (MCFL) phases within the framework of general relativity, we employ the ${\gamma}$-metric formalism. This formalism effectively incorporates deviations from spherical symmetry and allows a continuous transition between spherical (${\gamma=1}$) and axisymmetric (${\gamma\ne1}$) configurations. The corresponding spacetime geometry of the deformed compact object is described by the following line element (with ${G=c=1}$)~\cite{Zu17,Zub17,Terr21,Sahoo24}:
\begin{eqnarray}
{ds^2}=&&-e^{2{\nu(r)}}dt^2+\left(1-{\frac{2m(r)}{r}}\right)^{{-\gamma}}{dr^2} \nonumber\\
&&+{r^2 sin^2\theta d\phi^2}+{r^2d\theta^2}. 
\end{eqnarray}
where ${m(r)}$ denotes the gravitational mass enclosed within radius ${r}$, and ${\gamma}$ is the deformation parameter that quantifies the deviation from spherical symmetry. For ${\gamma<1}$, the star becomes oblate consistent with the effect of magnetic pressure being stronger in the perpendicular direction.\\
The stellar structure equations governing anisotropic magnetized quark stars can then be written as~\cite{Terr21,Sahoo24}:
\begin{eqnarray}
\label{eq:tov}
{\frac{dm}{dr}= 4\pi r^2 \gamma \epsilon_T}, \nonumber \\   
{\frac{dP_{\parallel T}}{dz}=-\frac{\left(\epsilon_T+P_{\parallel T}\right)\left[\frac{r}{2}+4\pi r^3 P_{\parallel T}-\frac{r}{2}\left(1-\frac{2m}{r}\right)^\gamma\right]}{\gamma r^2 \left(1-\frac{2m}{r}\right)^\gamma}}, \nonumber \\
{\frac{dP_{\perp T}}{dr}=  -\frac{\left(\epsilon_T+P_{\perp T}\right)\left[\frac{r}{2}+4  \pi r^3 P_{\perp T}-\frac{r}{2} \left(1-\frac{2m}{r}\right)^\gamma\right]}{ r^2 \left(1-\frac{2m}{r}\right)^\gamma}}. \nonumber \\
\end{eqnarray}
These equations are solved numerically in a way analogous to the standard Tolman-Oppenheimer-Volkoff (TOV) equations, but now using the anisotropic equation of state (EoS). Starting at the stellar center for a chosen central baryon number density ${n_{B0}=n_{B}(r=0)}$, one specifies the central energy density ${\epsilon_{T0}=\epsilon_{T}(r=0)}$, the central pressures ${P_{\parallel T0}=P_{\parallel T}(r=0)}$ and ${P_{\perp T0}=P_{\perp T}(r=0)}$, and defines the deformation parameter as
\begin{equation}
{\gamma=\frac{P_{\parallel T0}}{P_{\perp T0}}},
\end{equation}
which is taken to remain constant throughout the star.
The integration proceeds outward until the smaller of the two pressures (here ${P_{\parallel T}}$) vanishes, defining the polar radius ${Z}$ through ${P_{\parallel T}(Z)=0}$. The equatorial radius is then given by
\begin{equation}
{R=\frac{Z}{\gamma}},
\end{equation}
and the total gravitational mass is ${M=m(r=R)}$. Physically, the deformation parameter ${\gamma}$ encodes the influence of magnetic anisotropy: strong magnetic field (with ${P_{\perp T}>P_{\parallel T}}$) lead to smaller $\gamma$, yielding a more oblate configuration. When the magnetic field is switched off (${B=0}$), the pressures become isotropic (${P_{\parallel T}=P_{\perp T}}$) and ${\gamma=1}$, smoothly recovering the standard TOV equations for non-magnetized, spherically symmetric compact stars.
\begin{table*}[t]
\centering
\caption{Variation of the maximum mass ($M_{max}$) and corresponding equatorial radius ($R$) for different central magnetic field strengths (${B_{cen}}$) and bag constants (${B_{bag}}$) for both unpaired magnetized strange quark matter (MSQM) and paired magnetized color-flavor locked (MCFL) matter with pairing gaps $\Delta=\rm{30\,MeV}$ and $\rm{50\,MeV}$. The comparison highlights the influence of both the pairing gap and the magnetic field on the stellar structure.}
\setlength{\tabcolsep}{12pt}
\begin{tabular}{c|c|cc|cc|cc}
\toprule
\hline
${B_{cen}\, \rm{[G]}}$&${B_{bag}\,\rm{[MeV/fm^{3}]}}$&\multicolumn{2}{c|} {MSQM}& \multicolumn{2}{c|}{$\Delta = \rm{30 \, MeV\ (MCFL)}$} & \multicolumn{2}{c}{$\Delta =\rm{ 50 \, MeV\ (MCFL)}$}\\
\cmidrule(lr){3-4} \cmidrule(lr){5-6} \cmidrule(lr){7-8}
& &  $M_{max}$\,$\rm{[M_{\odot}]}$ &  $R$\,[$\rm{km}$] & $M_{max}$\,$\rm{[M_{\odot}]}$ & $R$\,[$\rm{km}$] &  $M_{max}$\,$\rm{[M_{\odot}}]$ &  $R$\,[$\rm{km}$] \\
\midrule
\hline
\multirow{2}{*}{0} & 65 & 1.73 & 9.63 & 1.83 & 10.17 & 2.06 & 11.08 \\
& 75 & 1.63    & 9.03   & 1.72    & 9.45    & 1.89     & 10.27    \\
\hline                                                                 
\multirow{2}{*}{$5 \times 10^{17}$} & 65 & 3.26  & 16.75 & 3.47 & 17.67 & 3.91 & 19.33 \\
& 75 & 2.97    & 15.70     & 3.19  & 16.39    & 3.54    & 17.75   \\   
\hline                               
\multirow{2}{*}{$7 \times 10^{17}$} & 65 & 3.47  & 16.95 & 3.71 & 17.70 & 4.14 & 19.45 \\
& 75 & 3.22    & 15.71     & 3.41  & 16.62   & 3.77   & 17.77   \\     
\hline                               
\multirow{2}{*}{$1 \times 10^{18}$} & 65 & 3.62  & 17.06 & 3.83 & 17.74 & 4.26 & 19.47 \\
& 75 & 3.36    & 15.80    & 3.55  & 16.70   & 3.91    & 17.84   \\     
\hline                                
\multirow{2}{*}{$1.5 \times 10^{18}$} & 65 & 3.68  & 17.06 & 3.89 & 17.75 & 4.32 & 19.49 \\
& 75 & 3.43    & 15.82    & 3.62 & 16.73    & 3.97    & 18.03   \\
\hline 
\bottomrule
\label{table1}
\end{tabular}
\end{table*}
\begin{figure*}[t]
  \centering
  \begin{subfigure}[b]{0.45\textwidth}
    \includegraphics[width=\linewidth]{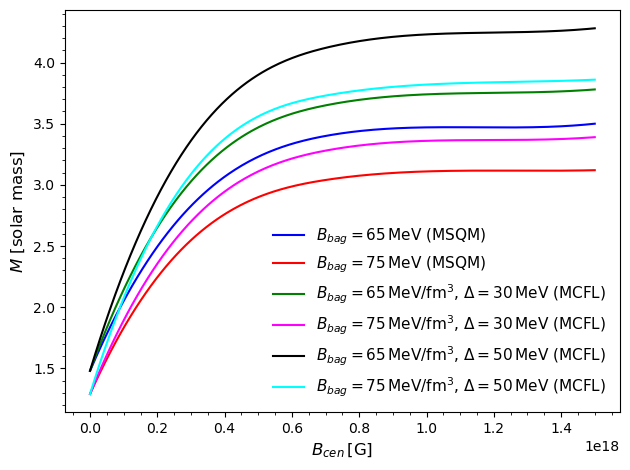}
  \end{subfigure}
  \hfill
  \begin{subfigure}[b]{0.45\textwidth}
    \includegraphics[width=\linewidth]{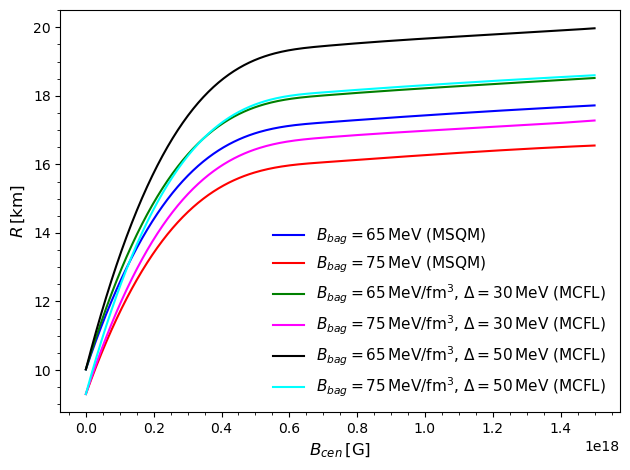}
  \end{subfigure}
  \vspace{0.5cm}  
\caption{ Variation of the stellar mass $M$ and corresponding equatorial radius $R$ with central magnetic field strength $B_{cen}$ for a fixed central baryon density $n_{B0}=\rm{0.5 \,fm^{-3}}$. The left panel shows the dependence of mass on $B_{cen}$, while the right panel presents the corresponding variation of equatorial radius. Results are shown for magnetized strange quark matter (MSQM) and magntized color-flavor locked (MCFL) phases with bag constants $B_{bag}=\rm{65\,MeV/fm^3}$ and $B_{bag}=\rm{75\,MeV/fm^3}$, and pairing gaps $\Delta=\rm{30\,MeV}$ and $\Delta=\rm{50\,MeV}$.}
\label{fig:MR_Bcen_PLOT}
\end{figure*}
\begin{figure*}[htbp]
  \centering
  \begin{subfigure}[b]{0.5\textwidth}
    \includegraphics[width=\linewidth]{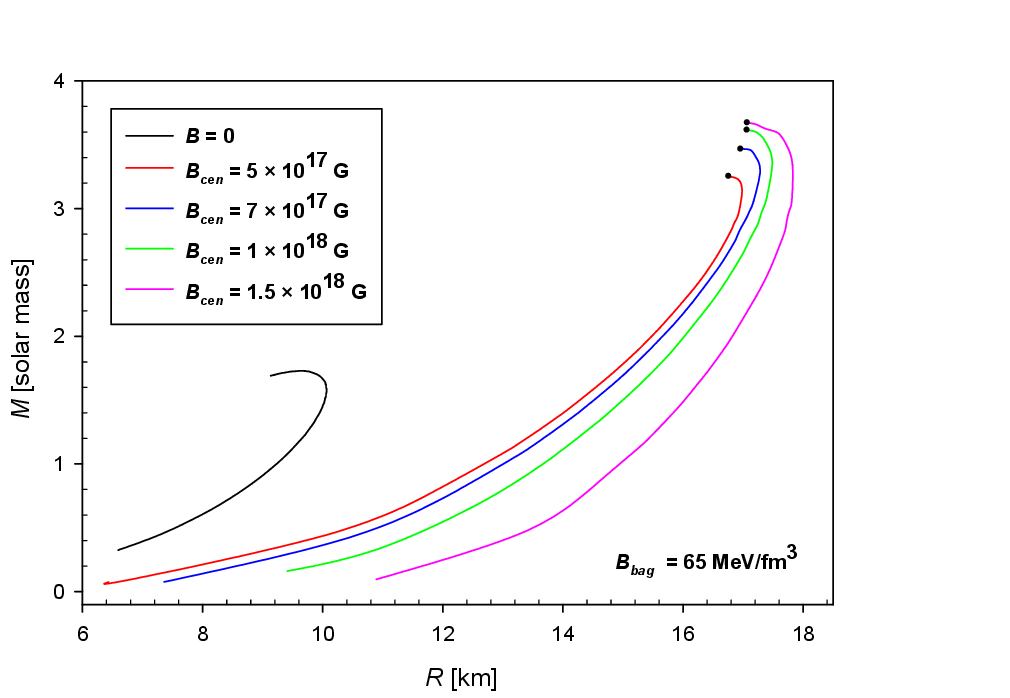}
  \end{subfigure}
  \hfill
  \begin{subfigure}[b]{0.45\textwidth}
    \includegraphics[width=\linewidth]{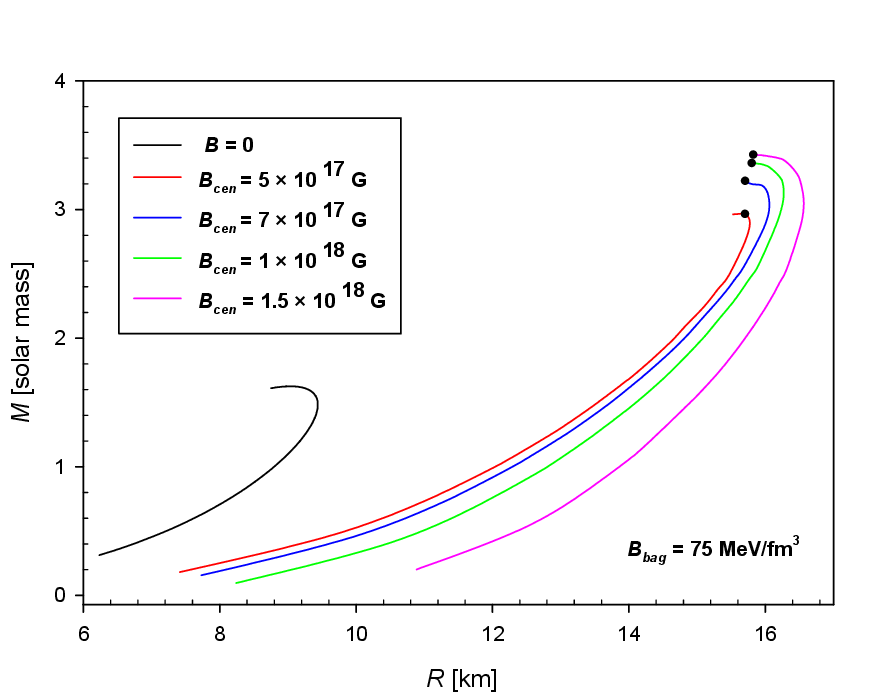}
  \end{subfigure}

  \vspace{0.5cm}  
  \begin{subfigure}[b]{0.45\textwidth}
    \includegraphics[width=\linewidth]{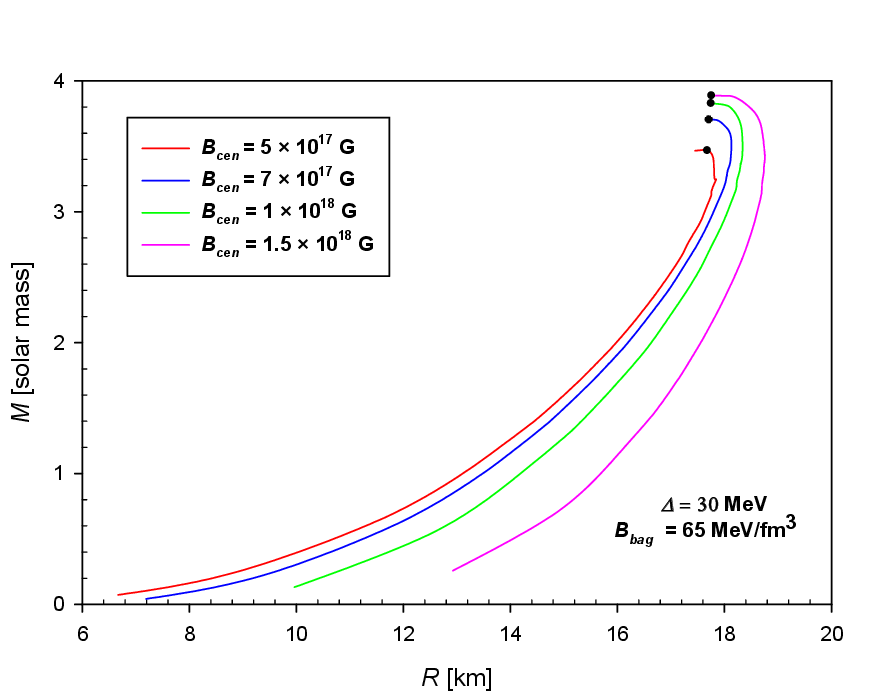}
  \end{subfigure}
  \hfill
  \begin{subfigure}[b]{0.45\textwidth}
    \includegraphics[width=\linewidth]{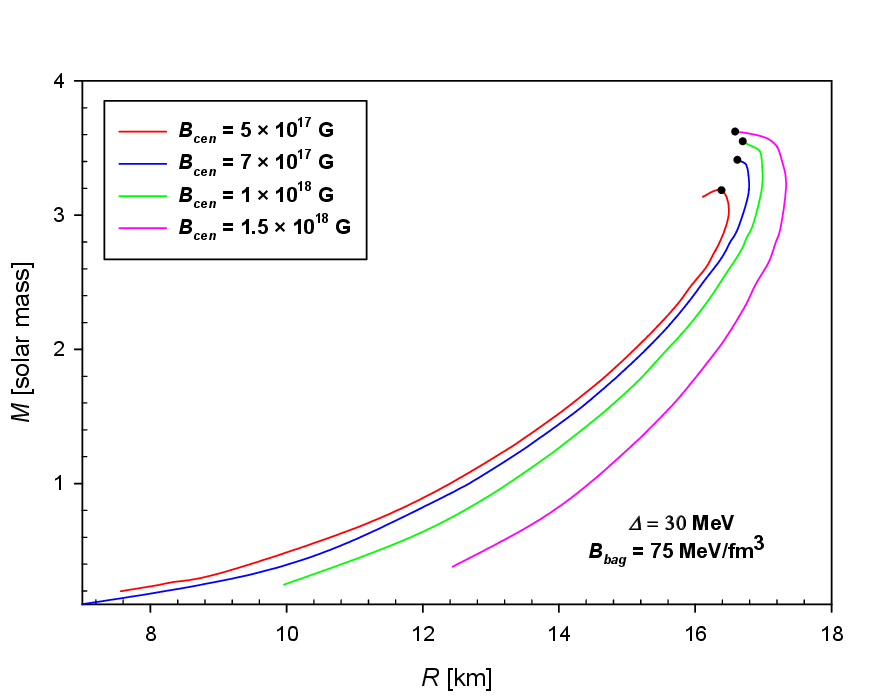}
  \end{subfigure}

  \vspace{0.5cm} 
  \begin{subfigure}[b]{0.45\textwidth}
    \includegraphics[width=\linewidth]{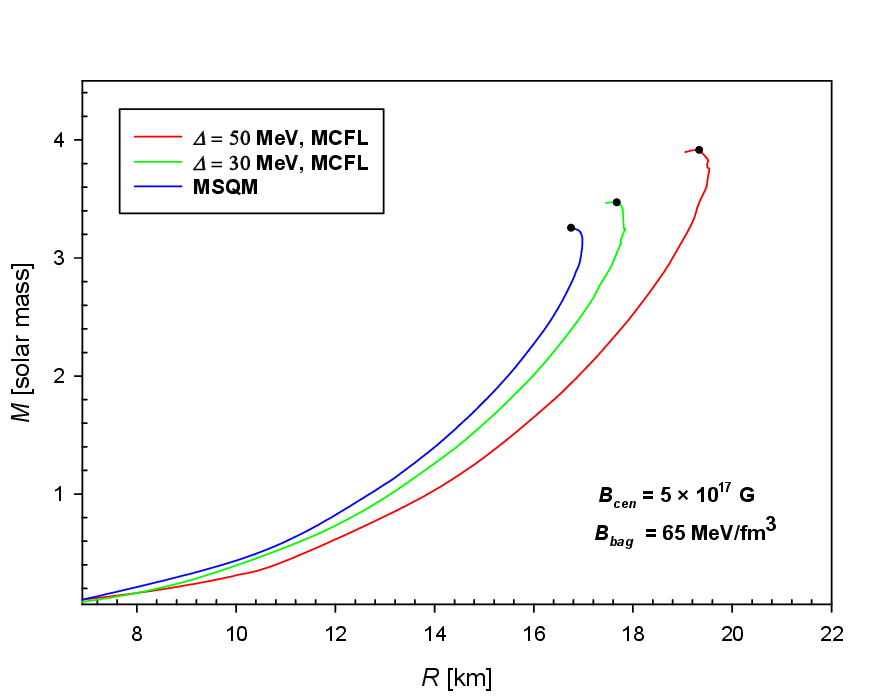}
    \label{fig:sub45}
  \end{subfigure}
  \hfill
  \begin{subfigure}[b]{0.45\textwidth}
    \includegraphics[width=\linewidth]{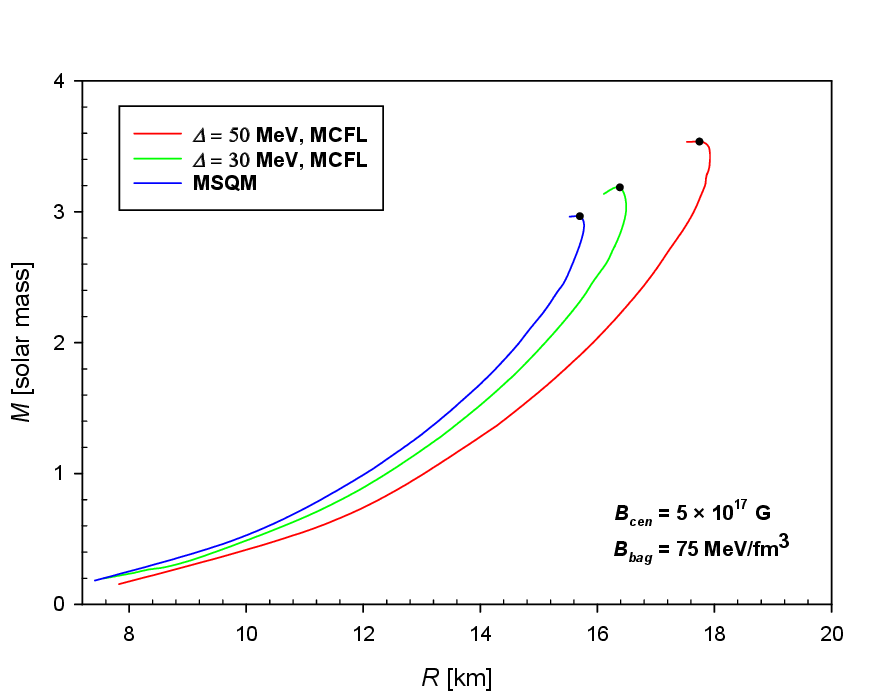}
  \end{subfigure}   
\caption{Variation of the stellar mass $M$ with equatorial radius $R$ for different central magnetic field strengths $B_{cen}=\rm{[5 \times 10^{17}, 7\times 10^{17}, 1\times 10^{18}, 1.5 \times 10^{18}]\,G}$. The top row shows the mass-radius relations for the magnetized strange quark matter (MSQM) phase with $B_{bag}=\rm{65\,MeV/fm^3}$ (left) and $B_{bag}=\rm{75\,MeV/fm^3}$ (right), together with the non-magnetized case ($B=0$). The middle row presents the corresponding results for the magnetized color-flavor locked (MCFL) phase for the same bag constants and a pairing gap of $\Delta=\rm{30\,MeV}$. The lower row compares the MSQM and MCFL phases. The dots indicate the maximum mass configurations in each case.}
\label{fig:mr_plot}
\end{figure*}
\section{Numerical Calculations Using Density-Dependent Magnetic Field}
\label{sec:5} 
As discussed in~\cite{Terr21,Sahoo24}, integrating the differential equations for the parallel and perpendicular pressures from Eqs.~\eqref{eq:tov} yields two corresponding total energy densities ${\epsilon_{\parallel T}}$ and ${\epsilon_{\perp T}}$, at each radial point. These are interpolated step by step during the numerical integration reflecting the anisotropic mass distribution within the magnetized quark star. To include this anisotropy consistently in the stellar structure equations, the total energy density is expressed as an average over the two components~\cite{Terr21,Sahoo24}:
\begin{eqnarray}
\label{eq:mtov} 
{\frac{dm}{dr}= 4\pi r^2 \gamma \left( \frac{\epsilon_{\parallel T}+ \epsilon_{\perp T}}{2}\right)}, \nonumber\\  
{\frac{dP_{\parallel T}}{dz}=-\frac{\left(\epsilon_{\parallel T}+P_{\parallel T}\right)\left[\frac{r}{2}+4\pi r^3 P_{\parallel T}-\frac{r}{2}\left(1-\frac{2m}{r}\right)^\gamma\right]}{\gamma r^2 \left(1-\frac{2m}{r}\right)^\gamma}}, \nonumber \\
{\frac{dP_{\perp T}}{dr}=  -\frac{\left(\epsilon_{\perp T}+P_{\perp T}\right)\left[\frac{r}{2}+4  \pi r^3 P_{\perp T}-\frac{r}{2} \left(1-\frac{2m}{r}\right)^\gamma\right]}{ r^2 \left(1-\frac{2m}{r}\right)^\gamma}}.\nonumber \\
\end{eqnarray}
Using the average energy density on the right hand side ensures that the mass density anisotropy is properly incorporated into the stellar configuration. Eqs.~\eqref{eq:mtov} are solved numerically with the required boundary conditions (discussed earlier) to determine the mass-radius relationship of deformed magnetized quark star.\\
{\bf{Magnetic Field Profile:}} In compact stars, the magnetic field is expected to vary with density, being stronger in the core and weaker near the surface. Instead of prescribing a purely density-dependent magnetic field ${B(n_{B})}$, we express it as a function of strange quark chemical potential ${\mu_s{}}$, which is a monotonic proxy for density at zero temperature (${\mu=E_{F}}$). This formulation provides numerical stability and better connects the microscopic quark properties to the macroscopic magnetic field distribution. The magnetic field profile is parameterized as~\cite{Ba97,Ba98,Sahoo24}:
\begin{equation} \label{eq:B}
{B_D \left(\mu_{s}\right) = B_S + B_0\left[1-exp\left(-
a\left(\frac{\mu_{s}}{\mu_{ 0s}}\right)^{b}\right)\right]},
\end{equation}
where ${B_{S}}$ (in units of ${B_{c}}$) is the surface magnetic field and ${\mu_{0s}}$ is the central chemical potential of strange quark. The constants ${a}$, ${b}$ and ${B_0}$ determine the shape and strength of the magnetic field profile. We take ${a=0.8}$ and ${b=0.9}$, ensuring a smooth nearly flat variation near both the center and the surface. The maximum central magnetic field is fixed at $\rm{1.5\times10^{18}\,G}$ consistent with the virial limit for magnetostatic stability. Since the baryon number density is non-zero at the stellar surface (${n_{B}\ne0}$), the above expression is modified to:
\begin{equation} \label{eq:BMod}
{B_D \left(\mu_{s}\right) = B_S + B_0\left[1-exp\left(-a\left(\frac{\mu_{s}-\mu_{s, min}}{\mu_{ 0 s}-\mu_{s,min}}\right)^{b}\right)\right]}.
\end{equation}
where ${\mu_{s,min}}$ corresponds to the strange quark chemical potential at the surface. This form ensures a consistent density-dependent magnetic field implemented via chemical potential smoothly connecting the surface and central field strengths in the deformed magnetized quark star.
\section{Results and Discussion}
\label{sec:6}
In this section, we investigate the influence of strong magnetic fields and color superconductivity on the structure and stability of magnetized quark stars comparing the magnetized strange quark matter (MSQM) phase with the magnetized color-flavor locked (MCFL) phase.
\subsection{\bf{Dependence of Stellar Mass and Radius on Central Magnetic Field}} 
To examine the influence of density-dependent magnetic fields on stellar structure, we explore the variation of the stellar mass and equatorial radius with the central magnetic field strength (${B_{cen}}$) at a fixed central baryon number density $n_{B0}=\rm{0.5\,fm^{-3}}$. As shown in the left panel of Fig.~\hyperref[fig:MR_Bcen_PLOT]{\ref*{fig:MR_Bcen_PLOT}}, the stellar mass increases rapidly with ${B_{cen}}$ at lower central magnetic field strengths and gradually saturated beyond $B_{cen}\sim\rm{10^{18}\,G}$. This behavior indicates that strong magnetic fields provide additional pressure support through magnetic anisotropy, effectively enhancing the gravitational mass. However, if the magnetic field exceeds a certain critical value, the star may lose equilibrium and undergo gravitational collapse. The magnetized color-favor locked (MCFL) phase yields higher masses compared to the magnetized strange quark matter (MSQM) phase reflecting the stabilizing influence of pairing interactions ($\Delta = \rm{30\,MeV}$ and $\rm{50\,MeV}$). Moreover, increasing the bag constant (${B_{bag}}$) leads to a decrease in the stellar mass consistent with a stiffer equation of state at lower ${B_{bag}}$. Similarly, the right panel of Fig.~\hyperref[fig:MR_Bcen_PLOT]{\ref*{fig:MR_Bcen_PLOT}} illustrate the variation of equatorial radius (${R}$) with ${B_{cen}}$. The radius follows a trend analogous to the mass, increasing with the central magnetic field due to the enhanced perpendicular pressure acting against gravity. The MCFL phase exhibits slightly larger radii than the MSQM phase for the same central magnetic field strength suggesting that quark pairing and strong magnetic field anisotropies jointly contribute to larger equilibrium configurations. These results collectively demonstrate that a density-dependent magnetic field significantly alters the mass-radius characteristics of quark stars leading to more massive and extended configurations under strong field conditions.
\subsection{\bf{Mass-Radius Relationship and Anisotropic Effects}} 
\begin{figure*}[t]
  \centering
  \begin{subfigure}[b]{0.44\textwidth}
    \includegraphics[width=\linewidth]{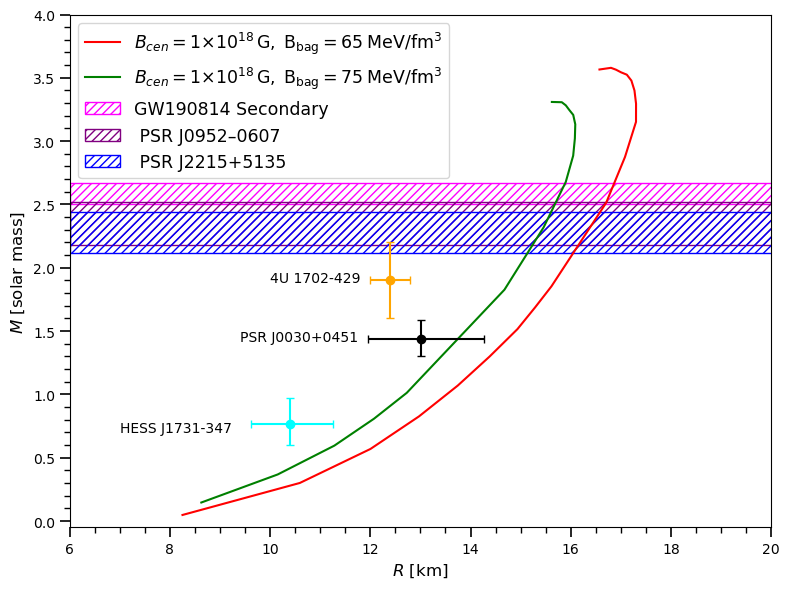}
  \end{subfigure}
  \hfill
  \begin{subfigure}[b]{0.45\textwidth}
    \includegraphics[width=\linewidth]{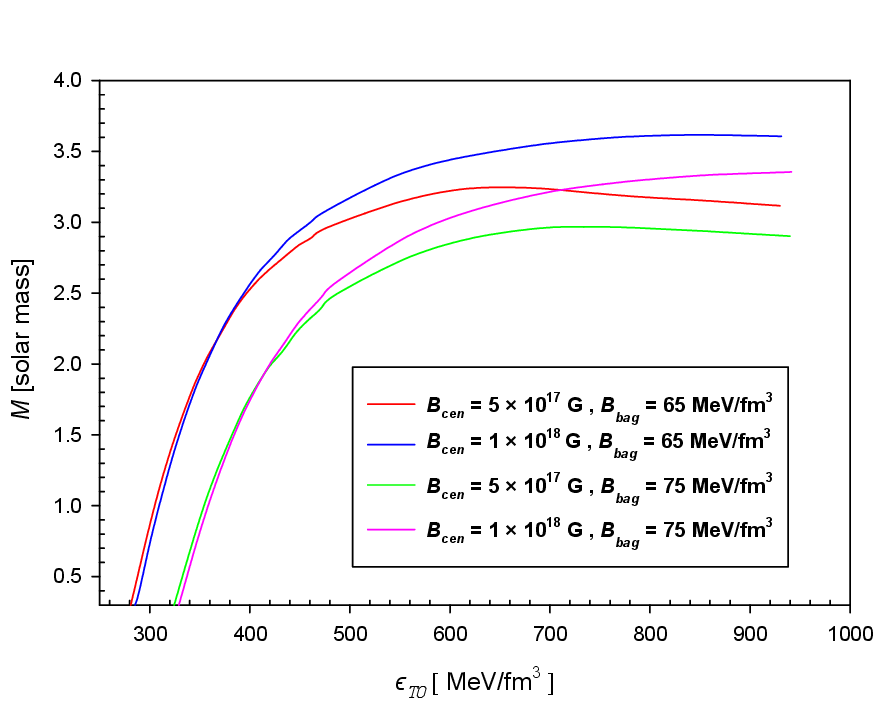}
  \end{subfigure}
\caption{ Variation of stellar mass with equatorial radius and with central total energy density for magnetized quark stars. The left panel shows the mass-radius relation at a central magnetic field strength $B_{cen}=\rm{1\times 10^{18}}\,G$ for two bag constants, $B_{bag}=\rm{65\,MeV/fm^3}$ (red curve) and $B_{bag}=\rm{75\,MeV/fm^3}$ (green curve). The shaded horizontal bands represent recent observational mass constraints from GW190814 secondary (pink)~\cite{Abbo20,Bis21}, $\rm{PSR\ J0952-0697}$ (magenta), and $\rm{PSR\ J2215+5135}$ (blue)~\cite{Rom22,Li18}. Recent mass-radius measurements from NICER and X-ray observations are also shown: $\rm{PSR\ J0030+0451}$ (black point with error bars)~\cite{Miller19}, $\rm{4U\ 1702-429}$ (orange point with error bars)~\cite{Natti17} and $\rm{HESS\ J1731-347}$ (cyan point with error bars)~\cite{Doro22}. These observational results indicate the range of viable stellar configurations in the light of current astrophysical measurements. The right panel shows the variation of mass with central total energy density $\epsilon_{T0}$ for magnetized quark stars at central magnetic field strengths $B_{cen}=\rm{[5\times 10^{17},1\times 10^{18}]\,G}$
for two values of the bag constant $B_{bag}=\rm{65\,MeV/fm^3}$ and $B_{bag}=\rm{75\,MeV/fm^3}$.}
  \label{fig:mr_compare}
\end{figure*}

\begin{figure*}[t]
  \centering
  \begin{subfigure}[b]{0.45\textwidth}
    \includegraphics[width=\linewidth]{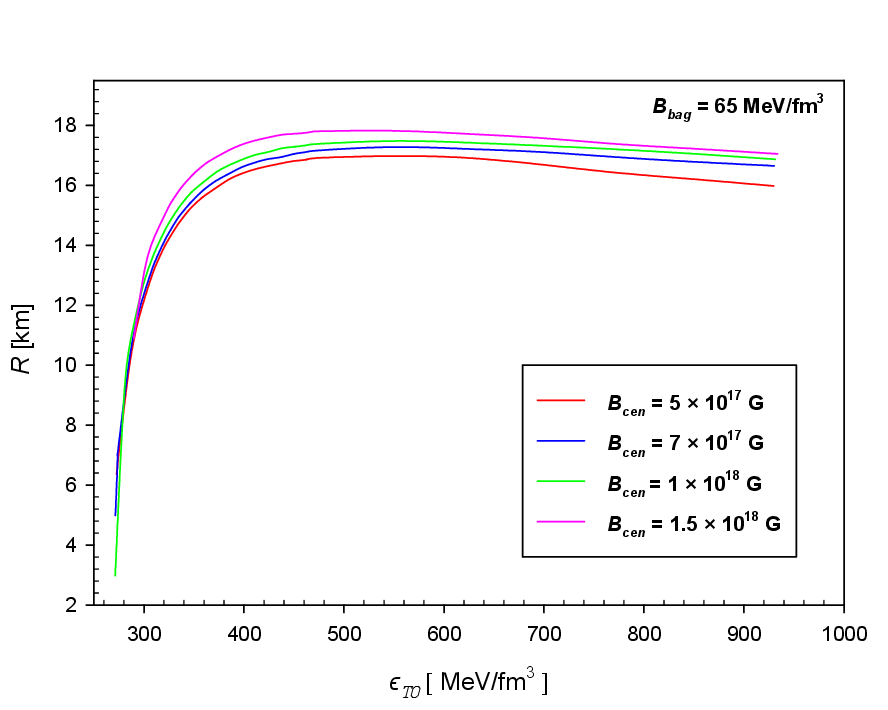}
  \end{subfigure}
  \hfill
  \begin{subfigure}[b]{0.45\textwidth}
    \includegraphics[width=\linewidth]{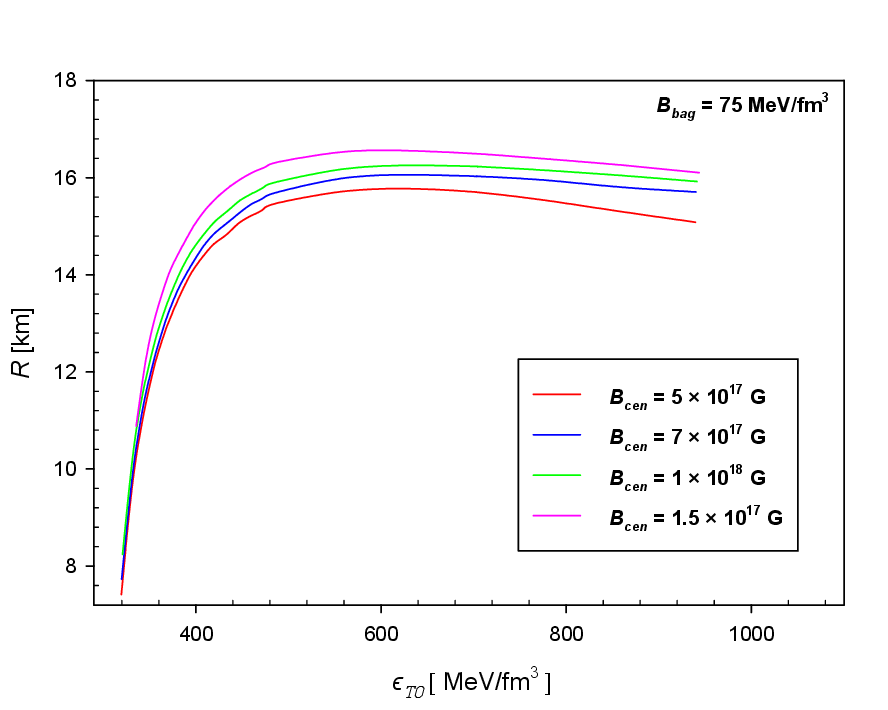}
  \end{subfigure}
  \vspace{0.5cm}  
\caption{ Variation of the equatorial radius with central total energy density for different central magnetic field strengths $B_{cen}=\rm{[5\times 10^{17}, 7\times 10^{17}, 1\times 10^{18}, 1.5\times 10^{18}]\,G}$. The left and right panels correspond to bag constants $B_{bag}=\rm{65\,MeV/fm^3}$ and $B_{bag}=\rm{75\,MeV/fm^3}$ respectively, showing the dependence of the equatorial radius $R$ on the central total energy density $\epsilon_{T0}$.}
  \label{fig:radius_plot}
\end{figure*}

The upper row panels of Fig.~\hyperref[fig:mr_plot]{\ref*{fig:mr_plot}}, illustrate the mass-radius relationship for quark stars composed of MSQM for two different bag constants $B_{bag}=\rm{65\,MeV/fm^3}$ and $\rm{75\,MeV/fm^3}$, at several central magnetic field strengths ${B_{cen}=[5\times10^{17}, 7\times10^{17}, 1\times10^{18}, 1.5\times10^{18}]\,G}$. For comparison, results for non-magnetized configurations are also included. For a fixed bag constant, both the maximum mass and the corresponding equatorial radius increase as the central magnetic field strength rises, as discussed earlier. For $B_{bag}=\rm{65\,MeV/fm^3}$, the maximum mass reaches $\rm{\sim 3.68\,M_{\odot}}$ with an equatorial radius of $\rm{\sim 17.06\ km}$ at $B_{cen}=\rm{1.5\times10^{18}\,G}$, while the corresponding non-magnetized configuration attains only $\rm{\sim 1.73\,M_{\odot}}$, with a radius of $\rm{\sim 9.63 \,km}$. This shows that strong magnetic fields significantly enhance the stellar mass and size. Physically, this occurs because the magnetic pressure counteracts gravitational collapse and the dominance of the total perpendicular pressure ${(P_{\perp T}>P_{\parallel T})}$ produces an oblate deformation (${R>Z}$), consistent with ${\gamma<1}$. The middle row panels of Fig.~\hyperref[fig:mr_plot]{\ref*{fig:mr_plot}} present the mass-radius relations for magnetized quark stars in the MCFL phase for the same bag constants and magnetic field strengths, considering pairing gaps $\Delta=\rm{30\,MeV}$ and $\rm{50\,MeV}$. The overall trend mirrors that of the MSQM case with stronger magnetic fields producing heavier and larger stars. For instance, at $\Delta=\rm{30\,MeV}$ and $B_{bag}=\rm{65 \, MeV/fm^3}$, the maximum mass is $\rm{\sim 3.89 \, M_{\odot}}$ with an equatorial radius of $\rm{\sim 17.75\,km}$, while for $\Delta=\rm{50\,MeV}$, the maximum mass rises further to $\rm{\sim 4.32\,M_{\odot}}$ with $\rm{R\sim 19.49\,km}$. Increasing the pairing gap leads to a stiffer EoS, as the color-flavor correlations reduce the system's compressibility, enhancing pressure support. When compared to the unpaired MSQM phase (lower row panels of Fig.~\hyperref[fig:mr_plot]{\ref*{fig:mr_plot}}), the MCFL configurations are noticeably more massive and larger, confirming that color superconductivity enhances the stability and maximum mass of magnetized quark stars. The left panel of Fig.~\hyperref[fig:mr_compare]{\ref*{fig:mr_compare}} shows that for a central magnetic field $B_{cen}=\rm{1\times10^{18}\,G}$, the maximum mass decreases as ${B_{bag}}$ increases from $\rm{65\,MeV/fm^3}$ to $\rm{75\,MeV/fm^3}$. A larger bag constant represents a stronger vacuum pressure, effectively softening the EoS and reducing the overall mass. We also include observational constraints from high mass pulsars and gravitational wave events. The magenta band in the left panel of Fig.~\hyperref[fig:mr_compare]{\ref*{fig:mr_compare}} denotes the secondary component of GW190814, with a measured mass of $M=\rm{2.59^{+0.08}_{-0.09}\; M_{\odot}}$~\cite{Abbo20,Bis21}. The purple and blue regions correspond to $\rm{PSR\ J0952-0607}$ ($M=\rm{ (2.35\pm0.17)\;M_{\odot}}$) and $\rm{PSR\ J2215+5135}$ ($M=\rm{2.27^{+0.17}_{-0.15}\; M_{\odot}}$)~\cite{Rom22,Li18}, respectively. Furthermore we consider additional constraints from precise X-ray pulse profile modeling and spectral analyses. The millisecond pulsar $\rm{PSR\ J0030+0451}$ observed by NICER provides simultaneous mass and radius measurements of $M=\rm{1.44_{-0.14}^{+0.15}\,M_{\odot}}$ and $R=\rm{13.02_{-1.06}^{+1.24}\,km}$~\cite{Miller19}. Similarly, the low mass X-ray binary $\rm{4U\ 1702-429}$ yields $M=\rm{1.9\pm 0.3\,M_{\odot}}$ and $R=\rm{12.4\pm 0.4 \,km}$~\cite{Natti17}, while the compact object $\rm{HESS\ J1731-347}$ is inferred to have $M=\rm{0.77_{-0.17}^{+0.20}\,M_{\odot}}$  and $R=\rm{10.4_{-0.78}^{+0.86}\,km}$~\cite{Doro22}. Our predicted configurations with $B_{cen}\ge \rm{5\times 10^{17}\,G}$ and moderate gap parameters are well within or above these observed mass ranges, suggesting that anisotropic magnetized quark stars could be viable candidates for these heavy compact objects. The right panel of Fig.~\hyperref[fig:mr_compare]{\ref*{fig:mr_compare}} and Fig.~\hyperref[fig:radius_plot]{\ref*{fig:radius_plot}}, display the dependence of total mass and equatorial radius on the total central energy density ${\epsilon_{T0}}$. As expected, for a given ${B_{bag}}$, both the mass and radius increase with the magnetic field strengths. For a central magnetic field, larger ${B_{bag}}$ values yield smaller, less massive configurations. The turning point in the ${M(\epsilon_{T0})}$ curve marks the onset of instability. Configurations satisfying ${\frac{\partial M}{\partial \epsilon_{T0}}>0}$ are stable against radial oscillations, while those beyond the maximum mass point with ${\frac{\partial M}{\partial \epsilon_{T0}}<0}$ becomes unstable. The observed behavior in the right panel of Fig.~\hyperref[fig:mr_compare]{\ref*{fig:mr_compare}} thus delineates the boundary between stable and unstable branches consistent with the general relativistic stability condition derived from the TOV formalism. We have listed the maximum masses ($M_{max}$) and corresponding equatorial radii ($R$) obtained for both the MSQM and MCFL phases with pairing gaps $\rm{\Delta=30\,MeV}$ and $\rm{50\,MeV}$, respectively in Table ~\hyperref[table1]{\ref*{table1}}. The results are presented for central magnetic field strengths $B_{cen}=\rm{[0,5\times10^{17},7\times10^{17},1\times10^{18},1.5\times10^{18}]\,G}$ and for bag constants $B_{bag}=\rm{65\,MeV/fm^3}$ and $\rm{75\,MeV/fm^3}$.
\begin{figure*}[t]
  \centering
  \begin{subfigure}[b]{0.45\textwidth}
    \includegraphics[width=\linewidth]{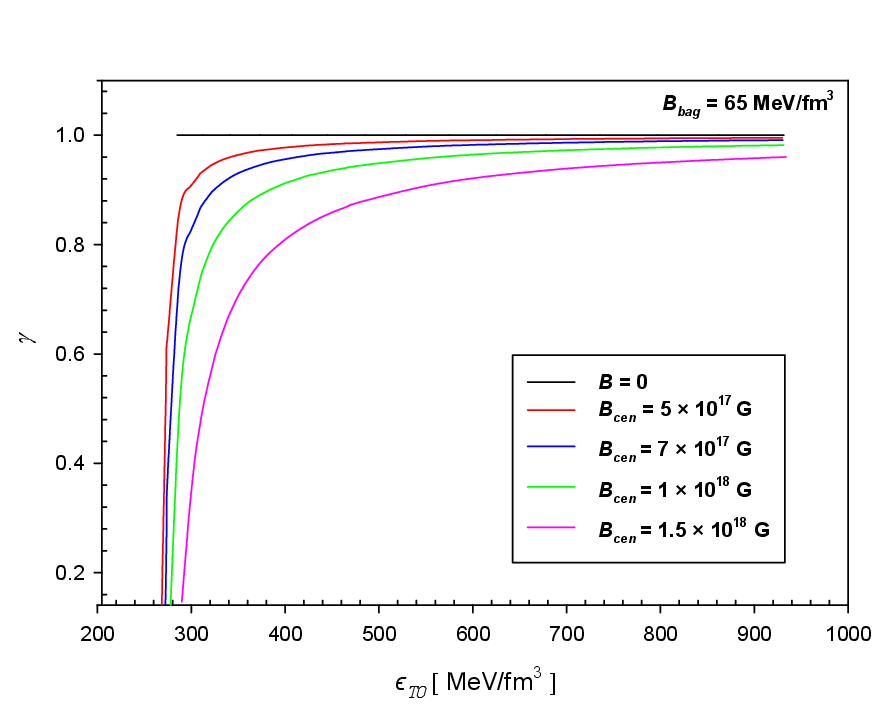}
  \end{subfigure}
  \hfill
  \begin{subfigure}[b]{0.45\textwidth}
    \includegraphics[width=\linewidth]{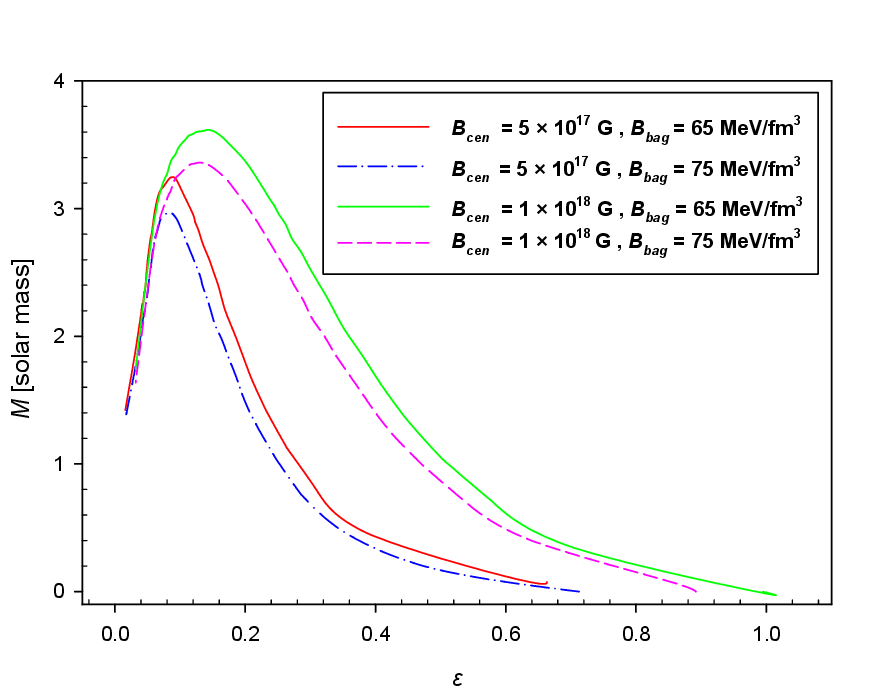}
  \end{subfigure}
\caption{ Magnetic field effects on the stability of magnetized quark stars. The left panel shows the variation of the anisotropy parameter $\gamma$ with central total energy density $\epsilon_{T0}$ for different central magnetic field strengths $B_{cen}=\rm{[0, 5 \times 10^{17}, 7\times 10^{17}, 1\times 10^{18}, 1.5 \times 10^{18}]\,G}$ at a fixed bag constant $B_{bag}=\rm{65\,MeV/fm^3}$. The right panel presents the dependence of gravitational mass $M$ (in solar mass) on ellipticity $\varepsilon$ for central magnetic field strengths $B_{cen}=\rm{[5\times 10^{17},1\times 10^{18}]\,G}$ and bag constants $B_{bag}=\rm{65\,MeV/fm^3}$ and $B_{bag}=\rm{75\,MeV/fm^3}$.}
  \label{fig:defor_plot}
\end{figure*}
\begin{figure*}[t]
  \centering
  \begin{subfigure}[b]{0.45\textwidth}
    \includegraphics[width=\linewidth]{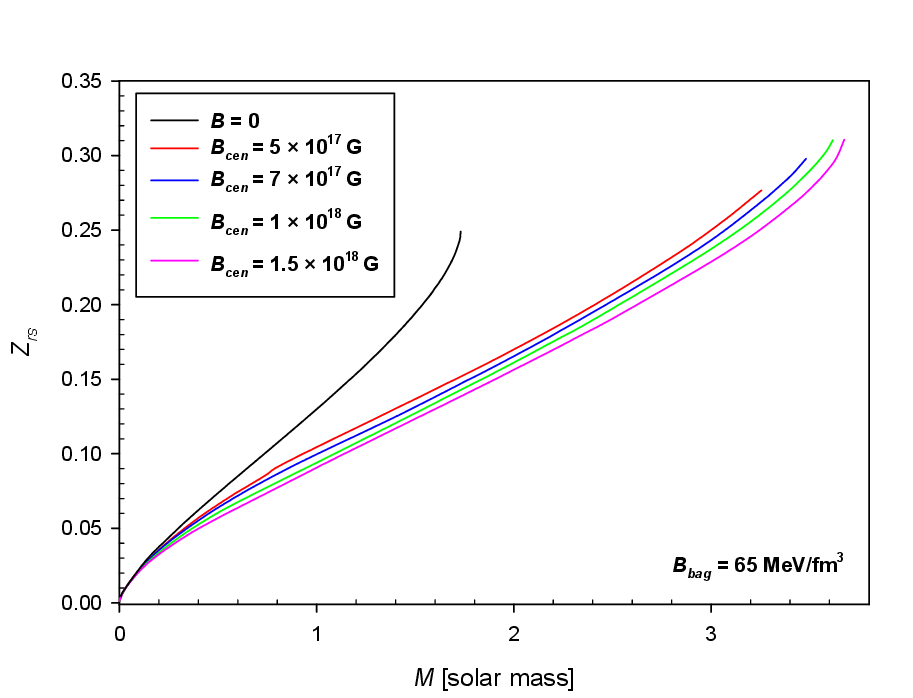}
  \end{subfigure}
  \hfill
  \begin{subfigure}[b]{0.45\textwidth}
    \includegraphics[width=\linewidth]{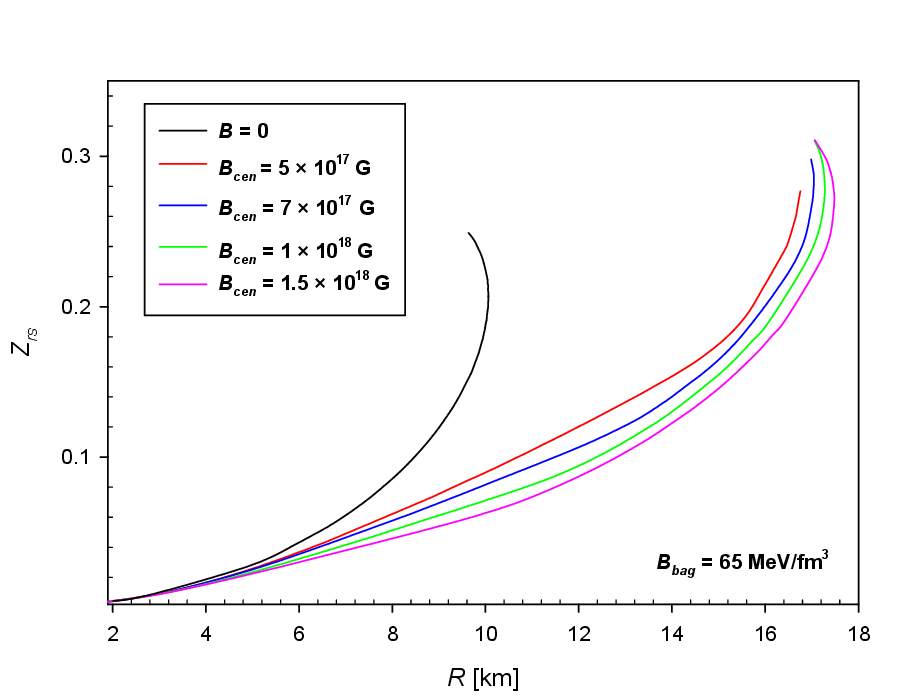}
  \end{subfigure}
  \vspace{0.5cm}  
  \begin{subfigure}[b]{0.45\textwidth}
    \includegraphics[width=\linewidth]{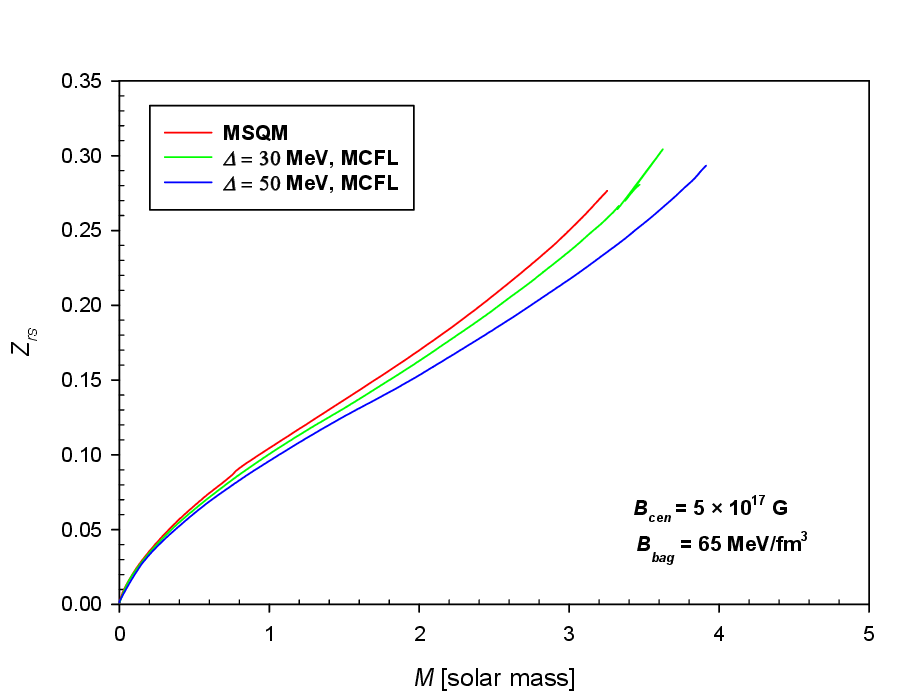}
  \end{subfigure}
  \hfill
  \begin{subfigure}[b]{0.45\textwidth}
    \includegraphics[width=\linewidth]{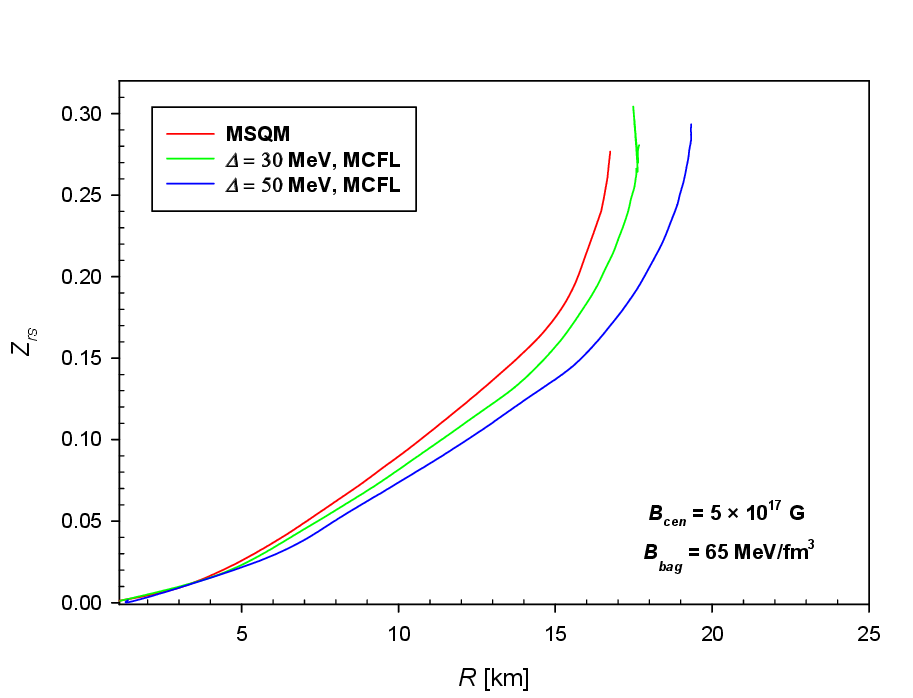}
  \end{subfigure}
\caption{ Gravitational redshift $Z_{rs}$ as a function of stellar mass and equatorial radius for magnetized quark stars under different physical conditions. The upper row panels correspond to the magnetized strange quark matter (MSQM) phase and show the effect of varying central magnetic field strengths $B_{cen}=\rm{[0, 5\times 10^{17},7\times10^{17},1\times 10^{18},1.5\times10^{18}]\,G}$ at a fixed bag constant $B_{bag}=\rm{65\,MeV/fm^3}$. The lower row panels correspond to the magnetized color-flavor locked (MCFL) phase and compare the results for pairing gaps $\Delta=\rm{30\,MeV}$ and $\Delta=\rm{50\,MeV}$ at a central magnetic field strength $B_{cen}=\rm{5\times 10^{17}\,G}$ and the same bag constant.}
  \label{fig:Grav_redshift_Plot}
\end{figure*}
\begin{figure*}[t]
  \centering
  \begin{subfigure}[b]{0.45\textwidth}
    \includegraphics[width=\linewidth]{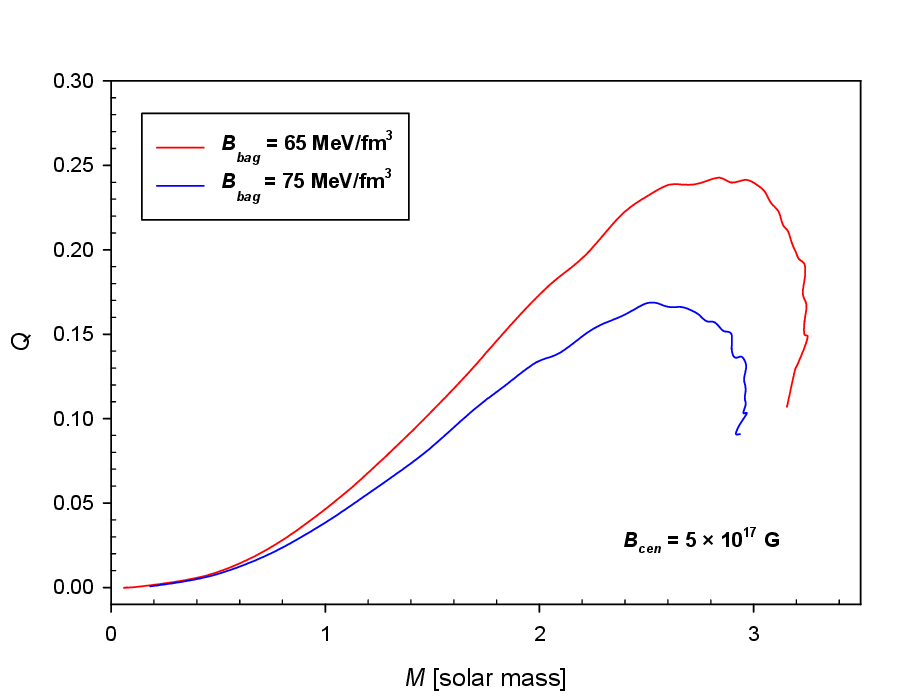}
  \end{subfigure}
  \hfill
  \begin{subfigure}[b]{0.45\textwidth}
    \includegraphics[width=\linewidth]{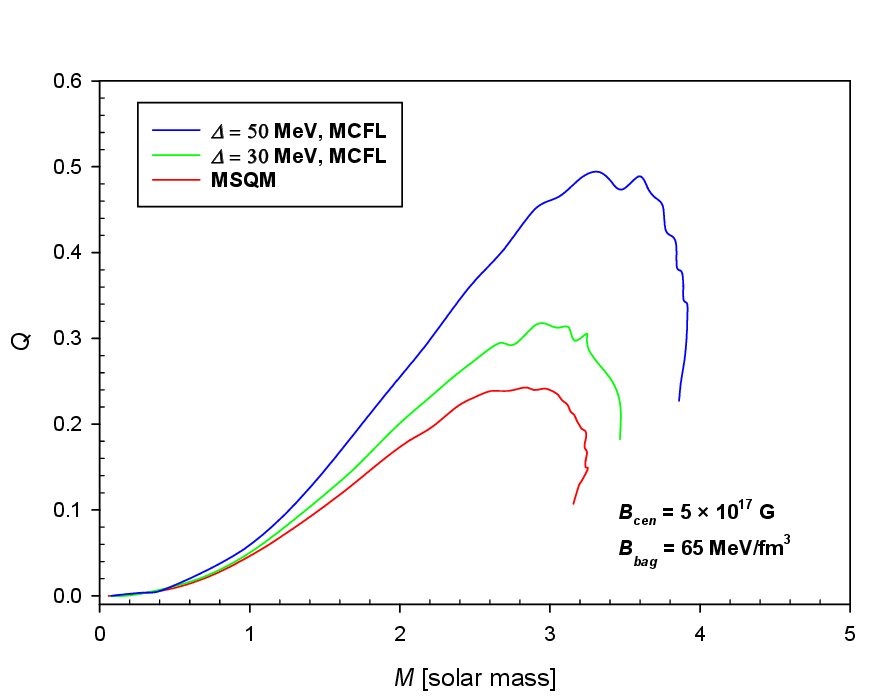}
  \end{subfigure}
\caption{ Variation of the quadrupole moment $Q$ with stellar mass $M$ for magnetized strange quark matter (MSQM) and magnetized color-flavor locked (MCFL) phases. The left panel shows $Q$ as a function of $M$ for the MSQM phase at fixed bag constants $B_{bag}=\rm{65\,MeV/fm^3}$ and $B_{bag}=\rm{75\,MeV/fm^3}$ for a central magnetic field strength $B_{cen}=\rm{5\times 10^{17}\,G}$. The right panel presents the corresponding variation for the MCFL phase with pairing gaps $\Delta=\rm{30\,MeV}$ and $\Delta=\rm{50\,MeV}$, compared with the MSQM case. All configurations in the right panel correspond to $B_{cen}=\rm{5\times 10^{17}\,G}$ and $B_{bag}=65\,MeV/fm^3$.}
\label{fig:Quadru_plot}
\end{figure*}
\subsection{\bf{Deformation and Ellipticity}}
The left panel of Fig.~\hyperref[fig:defor_plot]{\ref*{fig:defor_plot}} shows the variation of the deformation parameter (${\gamma}$-factor) with the central total energy density for different central magnetic field strengths, $B_{cen}= \rm{[0, 5\times 10^{17}, 7\times 10^{17},1\times 10^{18},1.5\times 10^{18}]\,G}$ at a fixed bag constant $B_{bag} =\rm{ 65 \, MeV/fm^3}$. As seen in the figure, the ${\gamma}$-factor decreases as the central magnetic field strength increases, indicating enhanced deformation of the stellar configuration. In the absence of magnetic field (${B=0}$), the star remains perfectly spherical, corresponding to ${\gamma=1}$. As the magnetic field grows stronger, magnetic pressure anisotropy leads to ${P_{\perp T}>P_{\parallel T}}$, producing an oblate deformation (${\gamma<1}$). The maximum deviation from spherical symmetry occurs at intermediate central energy densities, where the competing effects of magnetic pressure and matter pressure are most pronounced. At low central energy densities, the value of $\gamma$ is relatively small, indicating significant deformation, whereas at very high central energy densities, $\gamma$ tends toward unity, suggesting that the star becomes nearly spherical again. This behavior reflects a physical balance between gravitational compression and magnetic tension, with the magnetic contribution dominating only within a certain density range. To quantify the stellar deformation, we compute the ellipticity ${\varepsilon}$, defined as~\cite{Riz18}:
\begin{eqnarray}
{\varepsilon =\sqrt{1-\gamma}},
\end{eqnarray}
For a spherical configuration, ${\gamma \sim 1 }$, yielding ${\varepsilon \sim 0}$, while in the extreme deformation limit ${\gamma \to 0 }$, ${\varepsilon\to 1}$. Thus ${\varepsilon}$ serves as a direct measure of deviation from spherical symmetry. The right panel of Fig.~\hyperref[fig:defor_plot]{\ref*{fig:defor_plot}} presents the variation of stellar mass with ellipticity for two central magnetic field strengths $B_{cen}=\rm{[5\times10^{17},1\times 10^{18}]\,G}$, and for two bag constants $B_{bag}=\rm{65\,MeV/fm^3}$ and $\rm{75\,MeV/fm^3}$. At a central magnetic field $B_{cen}$, ellipticity decreases with increasing bag constant, implying that a stiffer vacuum pressure (higher ${B_{bag}}$) produces a more compact and less deformed configuration. Conversely, for a fixed bag constant, the ellipticity increases with magnetic field strength, demonstrating that stronger magnetic fields drive larger oblateness. Interestingly, the maximum stellar mass is achieved only for moderate ellipticity values, where the configuration is slightly deformed. Beyond this point, as the deformation increases, the maximum mass begins to decline, suggesting that highly deformed stars may become gravitationally unstable. This behavior indicates that strong magnetic fields, while capable of supporting higher masses up to a limit, can eventually induce structural instability when the anisotropy becomes excessive. This discussion pertain to the MSQM phase. The deformation and ellipticity in the MCFL phase exhibit similar qualitative behavior to those of the MSQM phase. However, their magnitudes are comparatively smaller because the quark pairing in the MCFL phase introduces an additional pressure component that stiffens the equation of state. This enhanced stiffness provides greater resistance to magnetic distortion, leading to reduced deformation and ellipticity for the same magnetic field strength. Although no separate plot is presented for MCFL phase, the results follow the same trend as discussed for the MSQM phase. Finally, it should be noted that the ${\gamma}$-metric formalism employed here remains accurate only for small to moderate deviations from spherical symmetry. For extreme field strengths or very high ellipticities (${\varepsilon\ge0.4}$), a full axisymmetric general relativistic treatment would be necessary to model the geometry and internal field structure of such highly deformed stars.
\subsection{\bf{Gravitational Redshift}} 
Gravitational redshift is one of the fundamental predictions of Einstein's General Theory of Relativity and serves as a key observational test of strong field gravity. It arises due to the influence of gravitational potential on the frequency of emitted radiation: photons escaping from the surface of a compact star lose energy while climbing out of the gravitational well, leading to an increase in their wavelength--a phenomenon known as redshift. For magnetized quark stars, which are extremely dense and possess intense magnetic fields, the gravitational redshift can become substantially modified, especially when the star is deformed under strong magnetic fields. The gravitational redshift for a spheroidal static configuration can be expressed as~\cite{Zubai15}:
\begin{equation}
\label{redshift}
{Z_{rs} = {\left(1-\frac{2M}{R}\right)^{-\gamma/2}} -1},
\end{equation}
where ${M}$ denotes the gravitational mass, ${R}$ is the equatorial radius, and ${\gamma}$ represents the deformation parameter quantifying the deviation from spherical symmetry. In the spherical limit (${\gamma=1}$), Eq. (\ref{redshift}) reduces to the standard redshift relation for isotropic compact star. The upper row panels of Fig.~\hyperref[fig:Grav_redshift_Plot]{\ref*{fig:Grav_redshift_Plot}} illustrate the variation of the gravitational redshift ${Z_{rs}}$ as a function of mass $\rm{M}$ and equatorial radius ${R}$ for magnetized strange quark matter (MSQM) phase under various central magnetic field strengths $B_{cen} = \rm{[0, 5\times 10^{17}, 7\times 10^{17},1\times 10^{18},1.5\times 10^{18}] \,G}$ at a fixed bag constant $B_{bag}=\rm{65\,MeV/fm^3}$. From both panels, it is evident that the redshift increases systematically with magnetic field strength. Deformed magnetized configurations thus exhibit higher surface redshifts compared to their non magnetized counterparts. This distinct trend provides a possible observational signature distinguishing magnetized strange quark stars from ordinary quark stars. The lower row panels of  Fig.~\hyperref[fig:Grav_redshift_Plot]{\ref*{fig:Grav_redshift_Plot}} compare the magnetized strange quark matter (MSQM) phase with the magnetized color-flavor locked (MCFL) phase for two different pairing gaps, $\Delta=\rm{30\,MeV}$ and $\rm{50\,MeV}$ at a fixed central magnetic field $B_{cen}=\rm{5\times 10^{17}\,G}$. The results demonstrate that increasing the pairing gap enhances the gravitational redshift, particularly for $\Delta=\rm{50\,MeV}$. This reflects the formation of a dense and more compact stellar configuration, consistent with the higher binding energy associated with stronger color superconducting gaps. A similar qualitative behavior is also observed for higher bag constant $B_{bag}=\rm{75\,MeV/fm^3}$, which is not shown here for brevity.
\begin{figure*}[t]
\centering
\begin{subfigure}[b]{0.45\textwidth}
\includegraphics[width=\linewidth]{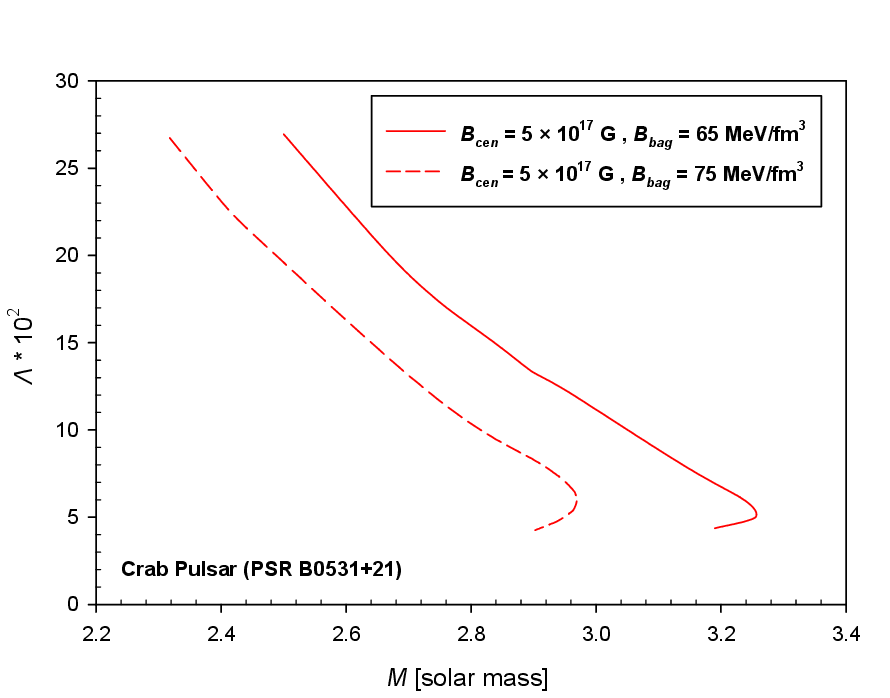}
\end{subfigure}
\hfill
\begin{subfigure}[b]{0.45\textwidth}
\includegraphics[width=\linewidth]{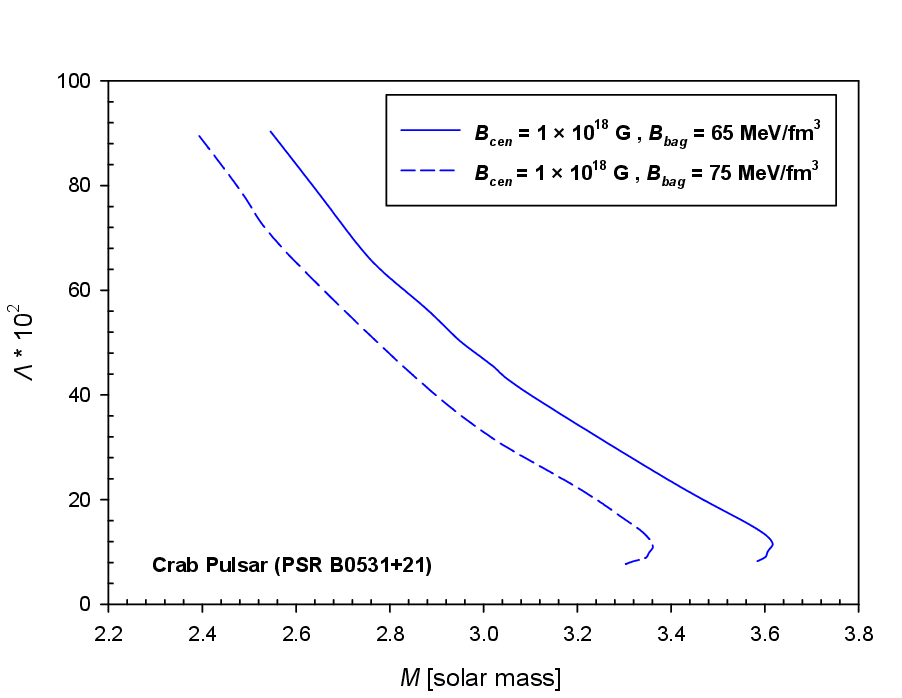}
\end{subfigure}
\vspace{0.5cm}  
\begin{subfigure}[b]{0.45\textwidth}
\includegraphics[width=\linewidth]{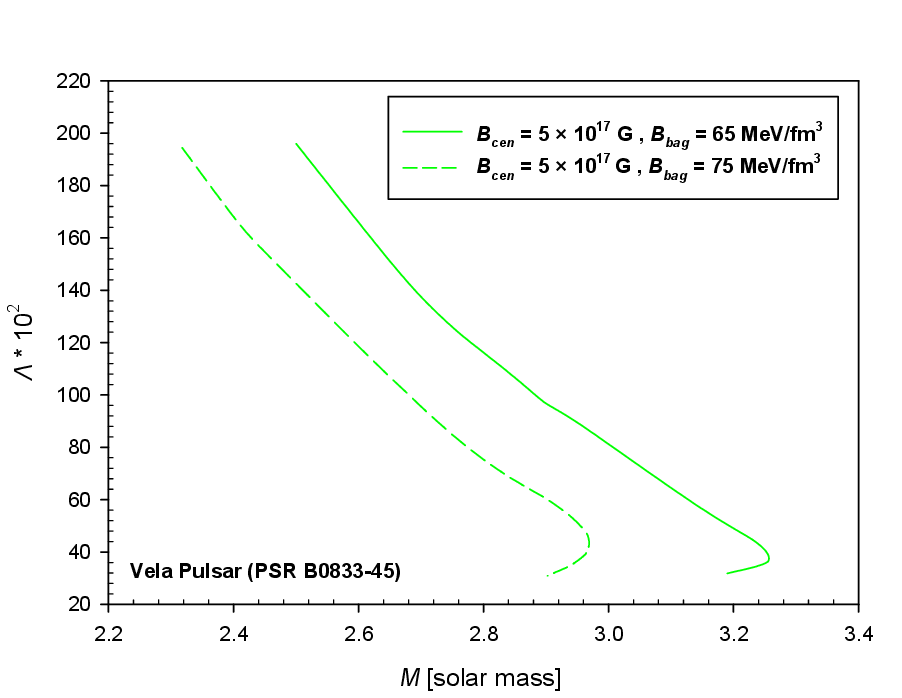}
\end{subfigure}
\hfill
\begin{subfigure}[b]{0.45\textwidth}
\includegraphics[width=\linewidth]{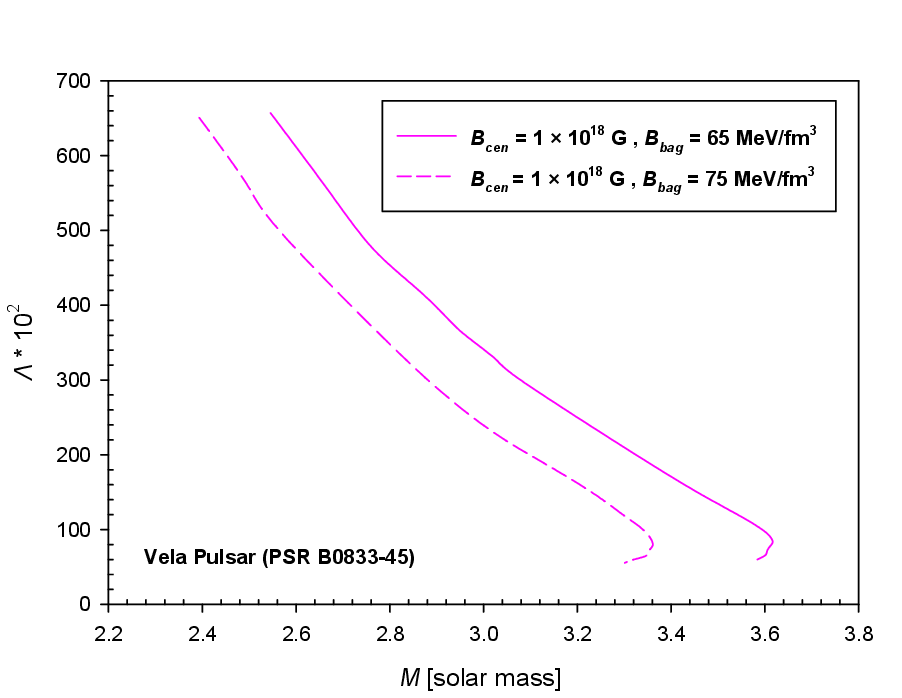}
\end{subfigure}
\caption{ Dimensionless tidal deformability $\Lambda$ as a function of stellar mass $M$ for the 
Crab Pulsar ($\rm{PSR\ B0531+21)}$ and the Vela Pulsar ($\rm{PSR\ B0833-45}$) under different magnetic field strengths and bag constants. The upper row panels correspond to $B_{cen}=\rm{5\times 10^{17}\,G}$, while the lower row panels represent $B_{cen}=\rm{1\times 10^{18}\,G}$. Each panel compares results for two bag constants,$B_{bag}=\rm{65\,MeV/fm^3}$ (solid lines) and $B_{bag}=\rm{75\,MeV/fm^3}$ (dashed lines).}
\label{fig:tidal_plot}
\end{figure*}
\subsection{\bf{Mass Quadrupole Moment}}
The mass quadrupole moment quantifies the degree to which a star's mass distribution deviates from perfect spherical symmetry. In deformed magnetized quark stars, strong magnetic fields induce pressure anisotropies that distort the stellar structure, resulting in a non zero quadrupole moment. Such deformations make these stars potential continuous sources of gravitational waves. The quadrupole moment for a spheroidal configuration can be expressed as~\cite{Her98}:
\begin{equation}
{Q = \frac{\gamma}{3} M^3 (1-\gamma^2)},
\label{eq:Q}
\end{equation}
where ${M}$ is the stellar mass and ${\gamma}$ denotes the deformation parameter. For spherically symmetric stars (${\gamma=1}$), the quadrupole moment vanishes (${Q=0}$), implying the absence of gravitational wave emission, as such radiation requires a time-varying quadrupole component. The left panel of Fig.~\hyperref[fig:Quadru_plot]{\ref*{fig:Quadru_plot}} shows the variation of the mass quadrupole moment ${Q}$ with stellar mass for deformed magnetized strange quark stars at a fixed central magnetic field $B_{cen}=\rm{5\times10^{17}\,G}$, considering two values of the bag constant, $B_{bag}=\rm{65\,MeV/fm^3}$ and $\rm{{75\,MeV/fm^3}}$. The results indicate that ${Q}$ decreases as ${B_{bag}}$ increases, suggesting that a softer equation of state (higher ${B_{bag}}$) leads to a less compact and less deformed configuration. The right panel of Fig.~\hyperref[fig:Quadru_plot]{\ref*{fig:Quadru_plot}} compares the results for magnetized strange quark matter (MSQM) and the magnetized color-flavor locked (MCFL) phase for two superconducting gaps, $\Delta=\rm{30\,MeV}$ and $\rm{50\,MeV}$. The quadrupole moment increases with the value of ${\Delta}$, indicating that stronger pairing in the MCFL phase enhances both the mass and deformation of the star. The MCFL configurations with $\Delta=\rm{50\,MeV}$ exhibit the largest ${Q}$ values, followed by those with $\Delta=\rm{30\,MeV}$, whereas the MSQM phase yields lower quadrupole moments and smaller maximum masses. The maximum ${Q}$ occurs for stars with intermediate mass and deformation i.e., those that are neither extremely massive nor highly  distorted. The small oscillatory features observed in the ${Q-M}$ curves arise from the quantized Landau levels induced by the strong magnetic field, which modulate the equation of state. Thus, the overall variation in the mass quadrupole moment is determined by the combined effect of stellar mass, deformation parameter ${\gamma}$, and internal pairing properties.
\subsection{\bf{Tidal Deformability}}
\begin{figure}[t]
  \vspace{0.0cm}
\eject\centerline{\epsfig{file=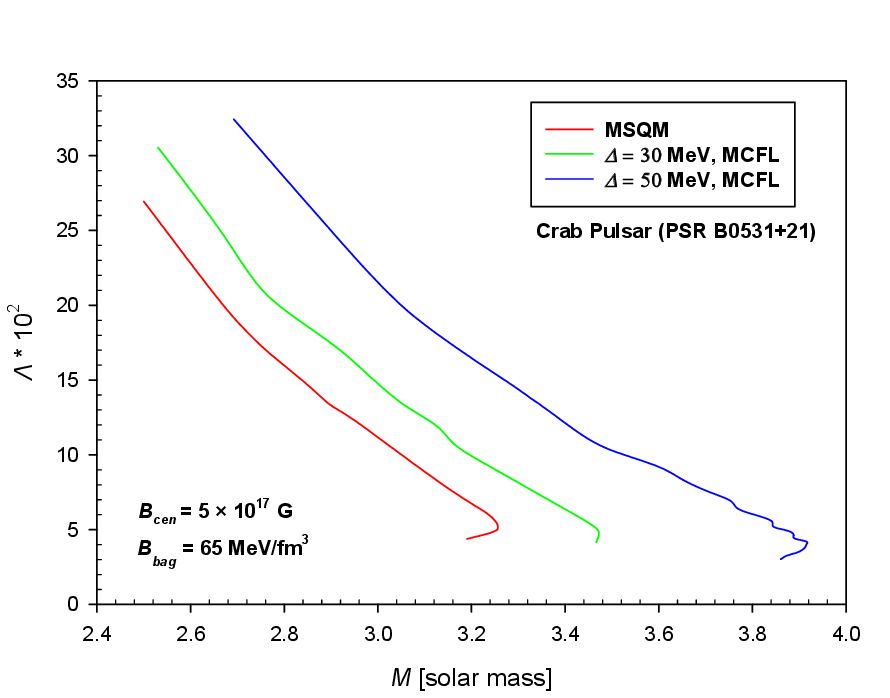,height=9cm,width=9cm}}
\caption{Variation of the dimensionless tidal deformability $\Lambda$ with stellar mass $M$ for the Crab Pulsar ($\rm{PSR\ B0531+21}$), considering different quark matter phases (MSQM and MCFL) at a central magnetic field strength $B_{cen}=\rm{5\times 10^{17}\,G}$ and a bag constant $B_{bag}=\rm{65\,MeV/fm^3}$.}
 \label{fig:Tidal-Compare}
\end{figure}

Tidal deformability characterizes how a compact star's shape responds to an external perturbation, such as rotation, gravitational interaction, or strong magnetic fields. In isolated magnetized pulsars, part of the rotational kinetic energy, ${E_{rot}=I\Omega^2}$, can be converted into electromagnetic or gravitational radiation. The resulting deformation is associated with a non-zero mass quadrupole moment ${Q}$, which depends on the magnetic field strength. In the Newtonian approximation, the external tidal field acting on the star is defined as ${\varepsilon_{tidal}=\Omega^2}$~\cite{P24,Mora04, Berti08}. The star's response to this field i.e., the induced quadrupole moment ${Q}$ determines its tidal deformability given by,
\begin{eqnarray}
\lambda=-\frac{Q}{\varepsilon_{tidal}}=-\frac{Q}{\Omega^2}=-\frac{\gamma}{12\pi^2}M^3(1-\gamma^2)P^2
\end{eqnarray}
where $\Omega=2\pi/P$. Using this formalism, we calculate the magnitude of tidal deformation for both magnetized strange quark matter (MSQM) and magnetized color-flavor locked (MCFL) configurations, employing the observed rotation periods of the Crab and Vela pulsars ($P_\text{{Crab}}=\rm{33\,ms}$, $P_\text{{Vela}}=\rm{89\,ms}$). The calculations are performed for central magnetic field strengths $B_{cen}=\rm{[5\times 10^{17}\,G}$ and $\rm{1\times10^{18}]\,G}$ and bag constants $B_{bag}=\rm{65\,MeV/fm^3}$ and $\rm{75\,MeV/fm^3}$. Fig.~\hyperref[fig:tidal_plot]{\ref*{fig:tidal_plot}} shows the variation of the dimensionless tidal deformability $\rm{\Lambda}$ as a function of stellar mass for two bag constants, $B_{bag}=\rm{65\,MeV/fm^3}$ and $\rm{75\,MeV/fm^3}$. From the upper row panels of  Fig.~\hyperref[fig:tidal_plot]{\ref*{fig:tidal_plot}}, we observe that ${\Lambda}$ increases with magnetic field strength, indicating that stronger fields enhance the star's susceptibility to deformation. However, at a central magnetic field strength $B_{cen}$, ${\Lambda}$ decreases with increasing ${B_{bag}}$, implying that softer equations of state (larger bag constants) produce more compact and less deformable configurations. The same qualitative trend is seen for the Vela pulsar in the lower row panels of Fig.~\hyperref[fig:tidal_plot]{\ref*{fig:tidal_plot}}, through the tidal deformability is slightly higher for the Vela pulsar under identical physical conditions. Since we use only the observed rotation period of the Crab and Vela pulsars in our analysis, the variation shown here corresponds to the MSQM phase under that rotational condition. Furthermore, Fig.~\hyperref[fig:Tidal-Compare]{\ref*{fig:Tidal-Compare}}, compares the dimensionless tidal deformability of magnetized strange quark matter (MSQM) with that of the magnetized color-flavor locked (MCFL) phase at $B_{cen}=\rm{5\times 10^{17}\,G}$ and $B_{bag}=\rm{65\,MeV/fm^3}$. In all cases, ${\Lambda}$ decreases monotonically with increasing stellar mass consistent with general relativistic expectations that more massive stars are more compact and less easily deformed. The MCFL stars with $\Delta=\rm{50\,MeV}$ exhibit the highest ${\Lambda}$ values, followed by $\Delta=\rm{30\,MeV}$, while MSQM stars show the lowest deformabilities. This behavior indicates that a larger superconducting gap in the MCFL phase corresponds to a stiffer equation of state and enhanced tidal deformability for a given stellar mass. Consequently, quark stars in the MCFL phase are more responsive to tidal perturbations than unpaired MSQM stars, a distinction that could have observable implications for gravitational wave signals from binary mergers.
\begin{figure*}[t]
\centering
\begin{subfigure}[b]{0.45\textwidth}
\includegraphics[width=\linewidth]{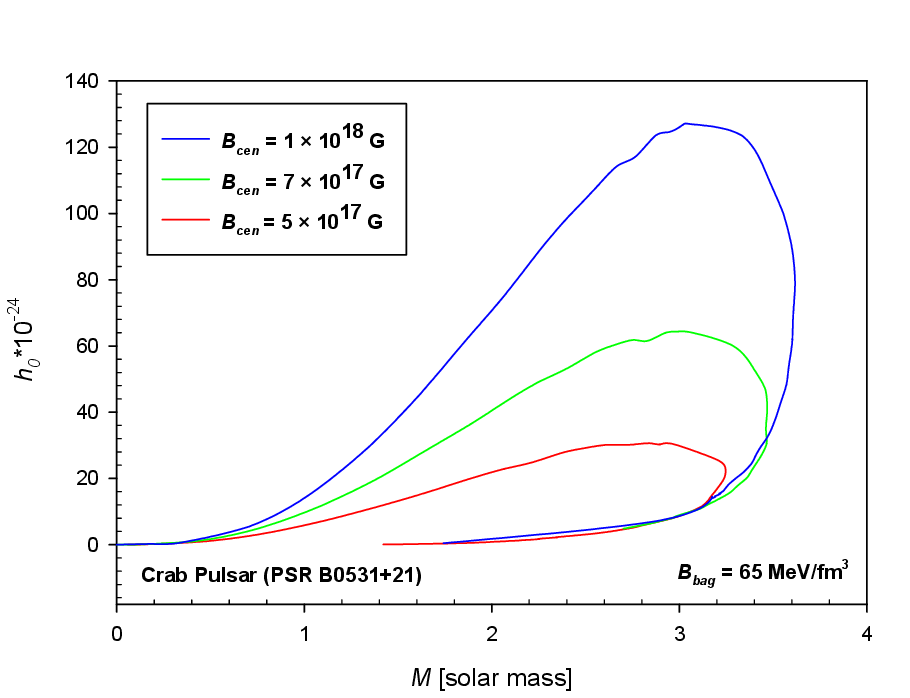}
\end{subfigure}
\hfill
\begin{subfigure}[b]{0.45\textwidth}
\includegraphics[width=\linewidth]{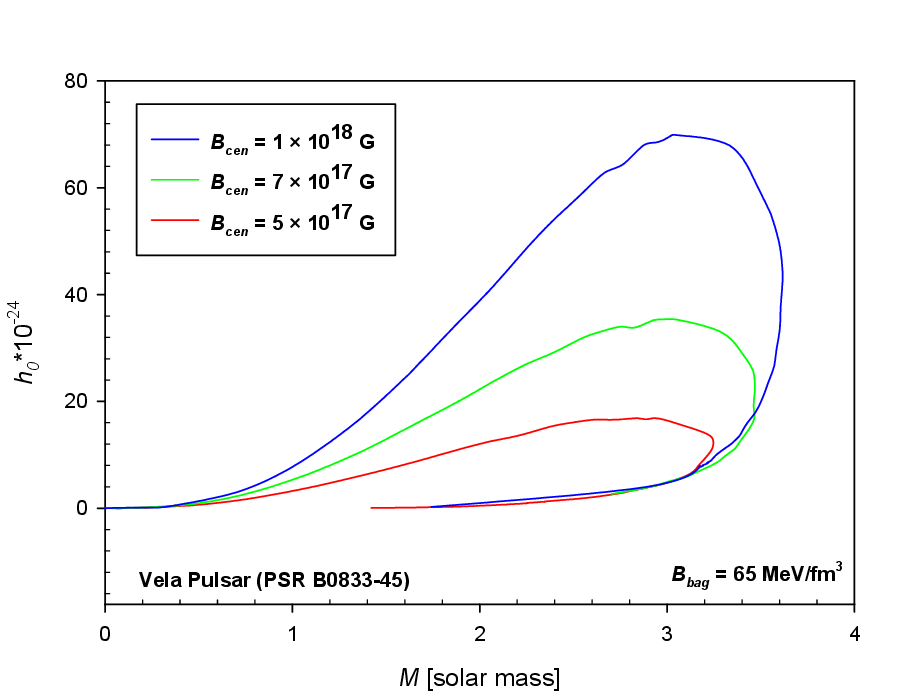} 
\end{subfigure}

\vspace{0.5cm}  

\begin{subfigure}[b]{0.45\textwidth}
\includegraphics[width=\linewidth]{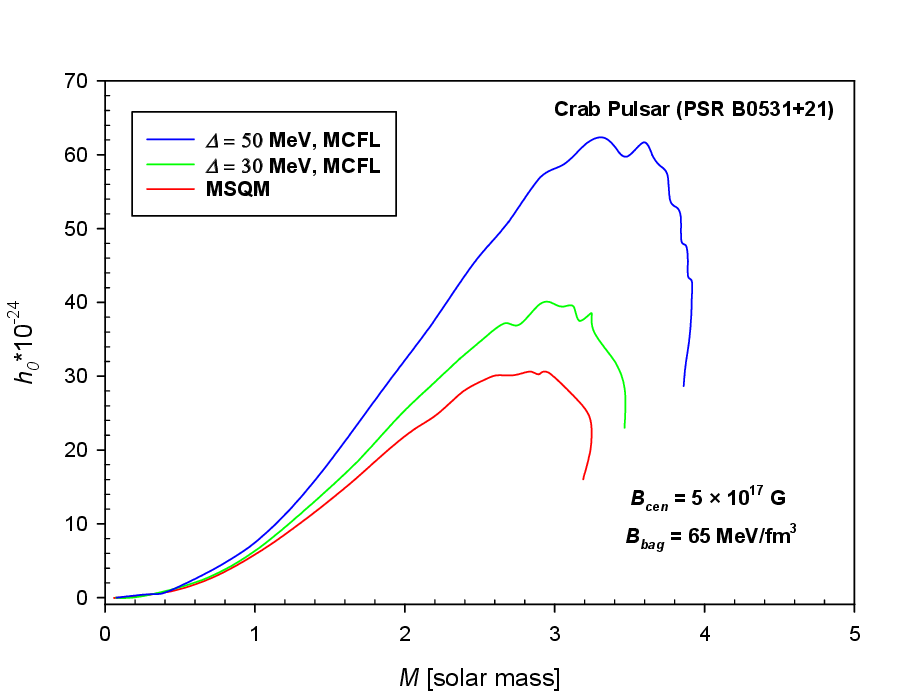}
\end{subfigure}
\hfill
\begin{subfigure}[b]{0.45\textwidth}
\includegraphics[width=\linewidth]{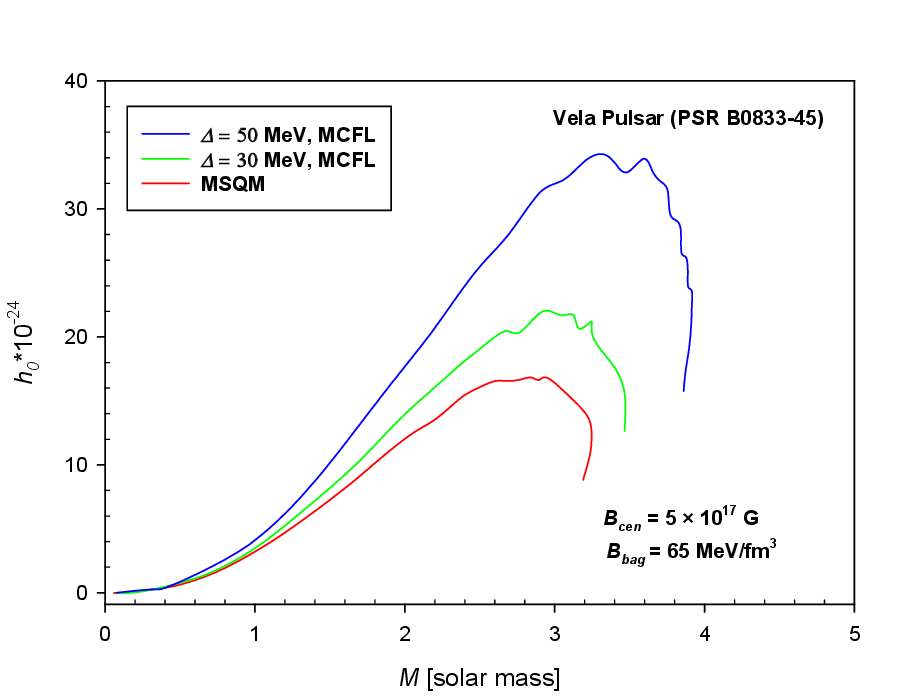}
\end{subfigure}

\caption{ Gravitational wave strain amplitude $h_{0}$ as a function of stellar mass $M$ for the 
Crab Pulsar $\rm{(PSR\ B0531+21)}$ and the Vela Pulsar $\rm{(PSR\ B0833-45)}$ under different magnetic field strengths and pairing gap parameters. The left panels show the variation of $h_{0}$ with $M$ for central magnetic field strengths $B_{cen}=\rm{[5\times 10^{17},1\times10^{17},1\times10^{18}]\,G}$ in the magnetized strange quark matter (MSQM) phase with $B_{bag}=\rm{65\,MeV/fm^3}$. The right panels illustrate the influence of pairing gaps $\Delta=\rm{30\,MeV}$ and $\Delta=\rm{50\,MeV}$ in the magnetized color-flavor locked (MCFL) phase, compared with the MSQM phase, for both pulsars at $B_{cen}=\rm{5\times 10^{17}\,G}$ and $B_{bag}=\rm{65\,MeV/fm^3}$.}
\label{fig:Grav_Compare}
\end{figure*}
\subsection{\bf{Gravitational Wave Amplitude}}
The gravitational wave amplitude or strain amplitude quantifies the fractional change in distance between two points caused by the passage of a gravitational wave. In the quadrupole approximation, it is expressed as~\cite{Flana05}:
\begin{equation}
{h_{0}=\frac{2G}{c^4}\frac{1}{r}\ddot{Q}},
\end{equation}
where ${\ddot{Q}}$ is the second time derivative of the mass quadrupole moment, ${G}$ is the gravitational constant, ${c}$ is the speed of light, and ${r}$ is the distance to the observer. This relation implies that gravitational waves are emitted by sources possessing a time varying, non zero mass quadrupole moment. A slightly deformed magnetized quark star naturally satisfies this condition, as the presence of strong magnetic field introduces pressure anisotropy that distorts the spherical symmetry and induces a quadrupole deformation. The strain amplitude for a rotating, slightly deformed star is given by~\cite{Bona96}:
\begin{equation}
{h_{0} = \frac{6G}{c^4}Q\frac{\Omega^2}{r}},
\end{equation}
where ${\Omega= {2\pi}/{P}}$ is the angular velocity and ${P}$ is the rotation period. Substituting the expression for ${Q}$ from Eq.~\eqref{eq:Q} and the angular velocity ${\Omega}$, we obtain the gravitational wave amplitude for a deformed magnetized quark star,
\begin{equation}
{h_{0}= 9.16 \; \gamma \;(1-\gamma^2)\left(\frac{M}{M_{\odot}}\right)^{3}\left(\frac{ms}{P}\right)^2\left(\frac{kpc}{r}\right)}.
\end{equation}
The strain amplitude ${h_{0}}$ is calculated for both the MSQM and MCFL phases using the observed parameters of two pulsars: the Crab pulsar ($P=\rm{33\,ms}$, $\dot P=\rm{4.2 \times 10^{-13}}$, $r=\rm{2\,kpc}$), and the Vela pulsar ($P=\rm{89\,ms}$, $\dot P=\rm{1.25 \times 10^{-13}}$, $r=\rm{0.5\,kpc}$). Fig.~\hyperref[fig:Grav_Compare]{\ref*{fig:Grav_Compare}} illustrate the variation of ${h_{0}}$ with the stellar mass for different central magnetic field strengths $B_{cen}= \rm{[5 \times 10^{17}, 7 \times 10^{17},1 \times 10^{18}]\,G}$ at a fixed bag constant $B_{bag}=\rm{65\, MeV/fm^3}$, considering both the MSQM and MCFL phases. From upper row panels of Fig.~\hyperref[fig:Grav_Compare]{\ref*{fig:Grav_Compare}}, the amplitude ${h_{0}}$ is found to increase with the central magnetic field strength. The enhancement in ${h_{0}}$ arises from the stronger magnetic anisotropy, which amplifies the quadrupole deformation of the star. The gravitational wave amplitude reaches its maximum for intermediate-mass configurations, where the magnetic distortion and self-gravity are optimally balanced. The lower row panels of Fig.~\hyperref[fig:Grav_Compare]{\ref*{fig:Grav_Compare}} show the variation of ${h_{0}}$ with stellar mass for both MSQM and MCFL phases at $B_{cen}=\rm{5\times10^{17}\,G}$. In all cases, ${h_{0}}$ increases with stellar mass, and the MCFL phase with gap parameter $\Lambda=\rm{50\,MeV}$ yields the largest amplitude. This indicates that the color superconducting MCFL phase, having a stiffer equation of state, allows for stronger magnetic deformation and consequently higher gravitational wave emission compared to the unpaired MSQM phase. Furthermore, for both the MSQM and MCFL phases, the amplitude corresponding to the Crab pulsar is nearly an order of magnitude higher than that of the Vela pulsar due to its shorter rotation period and hence larger angular velocity. The trend confirming that the interplay between magnetic field strength, rotation rate, and internal quark pairing govern the gravitational wave signal strength from such stellar sources. The computed gravitational wave strain amplitudes offer valuable insight into the astrophysical significance and observational prospects of magnetized quark stars as potential continuous gravitational wave emitters. Our results demonstrate that strong magnetic fields and the resulting pressure anisotropy can produce substantial stellar deformations, leading to strain amplitudes that approach or even enter the detectable range of current and next generation interferometers such as Advance LIGO, Virgo, and the Einstein Telescope. This finding implies that highly magnetized, rapidly rotating quark stars-particularly those in the MSQM and MCFL phases could contribute measurably to the persistent low-frequency gravitational wave background. Moreover, the dependence of the strain amplitude on the magnetic field strength and pairing gap parameters highlights the possibility of probing the internal composition and magnetic structure of quark stars through gravitational wave observations. However, these results not only complement electromagnetic constraints but also open a promising avenue for testing the existence and equation of state of magnetized quark matter via multimessenger astronomy.
\section{Summary and Conclusions}
\label{sec:7} 
In this work, we investigated the stellar structure and gravitational properties of magnetized spheroidal quark stars by considering two distinct phases of dense quark matter, magnetized strange quark matter(MSQM) and the magnetized color-flavor locked (MCFL) matter. Both were modeled using the MIT Bag model extended to include anisotropic pressures induced by strong magnetic fields. We adopted magnetic field profile, which varying with the chemical potential of quarks, providing a more physically motivated description of magnetic field stratification inside compact stars. The resulting anisotropy between the parallel and perpendicular pressure components leads to a departure from spherical symmetry, providing a spheroidal stellar configuration. To account for this deformation, we employed the $\gamma$-metric formalism, which modifies the Tolman-Oppenheimer-Volkoff (TOV) equations to include the effects of anisotropy and deformation. We explore the stellar configurations for a range of bag constants $B_{bag}=\rm{65\,MeV/fm^3}$ and $\rm{75\,MeV/fm^3}$, different central magnetic field strengths $B_{cen}=\rm{[0,5\times10^{17},7\times 10^{17},1\times 10^{18},1.5\times 10^{18}]\,G}$, and pairing gaps $\Delta=\rm{30\,MeV}$ and $\rm{50\,MeV}$ for the MCFL phase. The results demonstrate that the dependence of stellar properties on the central magnetic field strength ${B_{cen}}$ shows that both the stellar mass and equatorial radius increase with rising magnetic field intensity at a fixed central baryon number density $n_{B0}=\rm{0.5\,fm^{-3}}$. This behavior can be attributed to the enhancement of perpendicular pressure in the presence of strong magnetic fields, which counteracts gravitational compression and results in the overall expansion of the star. The increase becomes less pronounced beyond a certain field strength, indicating a saturation effect in magnetic support; however, if the magnetic field exceeds a critical value, the star becomes unstable and may eventually collapse. Moreover, the MCFL phase exhibits higher mass and radius values than the MSQM phase, reflecting the stiffening of the equation of state due to quark pairing. Both the maximum mass and equatorial radius increase with increasing magnetic field strength, whereas a larger bag constant softens the equation of state (EoS), yielding more compact stars. Due to quark pairing, the MCFL phase exhibits a stiffer EoS compared to MSQM, resulting in larger maximum masses and radii. Stability analysis based on ${\partial M/\partial \epsilon_{T0}>0}$ confirms that stars up to the maximum mass are stable against radial oscillations. We also examined the absolute stability of magnetized quark matter by comparing the energy per baryon $\left(E/A\right)$ at zero pressure with that of ${^{56}{Fe}}$, $(E/A)_{^{56}{Fe}}\simeq \rm{930\,MeV}$. For the parameter sets considered, both MSQM and MCFL matter with lower bag constant ${B_{bag}=65\,MeV/fm^3}$ and substantial pairing gaps $\Delta\ge\rm{30-50\,MeV}$ yield $(E/A)_{P=0}<\rm{930\,MeV}$, demonstrating absolute stability against hadronization. Higher bag constant $B_{bag}=\rm{75\,MeV/fm^3}$ leads to metastable configurations. Moreover, strong magnetic fields lower the energy per baryon through Landau quantization effects, further stabilizing magnetized quark matter in both phases. We further analyzed several key observables linked to deformation and magnetization: gravitational redshift, tidal deformability, mass quadrupole moment, and gravitational wave amplitude. The gravitational redshift increases with the central magnetic field, indicating stronger spacetime curvature near the stellar surface. The tidal deformability and quadrupole moment both grow with magnetic field strength, showing that enhanced anisotropy leads to greater stellar deformation. For a fixed magnetic field and bag constant, the quadrupole moment peaks for intermediate-mass stars and decreases with increasing bag constant. Our analysis of the gravitational wave strain amplitude (${h_{0}}$) shows values of the order of $\rm{10^{-24}}$, increasing with magnetic field strength and attaining maximum values for intermediate-mass configurations. Notably, MCFL stars with larger pairing gaps ($\Delta=\rm{50\,MeV}$) exhibit higher quadrupole moments and strain amplitudes than their MSQM counterparts, making them stronger potential emitters of continuous gravitational waves. These results underline the significant role of magnetic field anisotropy, bag constant and quark pairing in shaping the macroscopic observables of quark stars. In particular, the predicted gravitational wave amplitudes and tidal deformabilities fall within the sensitivity ranges of future detectors such as the Einstein Telescope (ET) and Cosmic Explorer (CE), while the high redshift and mass-radius predictions can be tested by NICER and upcoming X-ray missions. Overall, this study provides a self-consistent framework connecting microphysical modelling of magnetized quark matter with observable astrophysical signatures. Future work could extend this approach to include rotation, finite temperature effects, and color-magnetic interactions, potentially refining predictions for next generation gravitational wave and X-ray observations.
\section{Acknowledgments}
The authors acknowledge the National Supercomputing Mission (NSM) for providing computing resources of PARAM Porul at NIT Tiruchirappalli, implemented by C-DAC and supported by the Ministry of Electronics and Information Technology (MeitY) and the Department of Science and Technology (DST), Government of India. Rajasmita Sahoo sincerely thanks Mrutunjaya Bhuyan for carefully going through the manuscript and for his valuable suggestions.
\bibliography{reference} 
\bibliographystyle{spphys}




%
%


\end{document}